\documentclass[lettersize,journal,compsoc]{IEEEtran}
\usepackage{amsmath,amsfonts}
\usepackage{algorithmic}
\usepackage{array}
\usepackage[caption=false,font=normalsize,labelfont=sf,textfont=sf]{subfig}
\usepackage{textcomp}
\usepackage{stfloats}
\usepackage{url}
\usepackage{verbatim}
\usepackage{graphicx}
\usepackage{cite}
\usepackage{threeparttable}
\usepackage{hyperref}
\usepackage{url}

\usepackage{amsmath,amssymb,amsfonts}
\usepackage{textcomp}
\usepackage{xcolor}
\usepackage{subfig}
\usepackage{amsthm}
\usepackage{caption}
\usepackage{booktabs}
\usepackage{cleveref}
\usepackage{balance}
\theoremstyle{definition}

\usepackage{csquotes}
\usepackage{tikz}
\usepackage{paralist}
\usepackage{xspace}
\usepackage{enumitem}
\usepackage[linesnumbered,lined,boxed,commentsnumbered]{algorithm2e}
\usepackage{multirow}
\usepackage{color}
\definecolor{changeColor}{rgb}{0,0,255} 
\usepackage{marginnote}
\usepackage{colortbl}
\usepackage[most]{tcolorbox}

\definecolor{mygray}{RGB}{217, 217, 217}
\usepackage{makecell}
\newcommand{\tabitem}{~~\llap{\textbullet}~~}

\definecolor{myred1}{RGB}{248, 113, 103}

\newcommand{\redtext}[2]{{\color{myred1}\marginnote{#1}#2}}

\AtBeginDocument{%
  \providecommand\BibTeX{{%
    \normalfont B\kern-0.5em{\scshape i\kern-0.25em b}\kern-0.8em\TeX}}}

\newcommand{\ie}{\textit{i}.\textit{e}.\xspace}
\newcommand{\eg}{\textit{e}.\textit{g}.,\xspace}
\newcommand{\aka}{\textit{a.k.a.,}\xspace}

\AtBeginDocument{%
  \providecommand\BibTeX{{%
    \normalfont B\kern-0.5em{\scshape i\kern-0.25em b}\kern-0.8em\TeX}}}

\begin{document}

\title{What Characterizes Pairwise Modular Smells?}

\author{Chenxing~Zhong, Daniel Feitosa, Paris Avgeriou, Huang~Huang, Wei~Song, and He~Zhang \\

\IEEEcompsocitemizethanks{\IEEEcompsocthanksitem Chenxing Zhong and Wei Song are with the Nanjing University of Science and Technology, Nanjing 210094, China.\protect\\
Email: chenxingzhong@njust.edu.cn; wsong@njust.edu.cn
\IEEEcompsocthanksitem Daniel Feitosa and Paris Avgeriou are with the University of Groningen, Groningen 9700AB, the Netherlands.\protect\\
Email: d.feitosa@rug.nl; p.avgeriou@rug.nl
\IEEEcompsocthanksitem Huang Huang is with State Grid Nanjing Power Supply Company, Nanjing 210000, China.\protect\\
Email: sgcc.huang.huang@gmail.com
\IEEEcompsocthanksitem He Zhang is with the State Key Laboratory for Novel Software Technology, Nanjing University, Nanjing 210093, China.\protect\\
E-mail: 
hezhang@nju.edu.cn \\
}
\thanks{Manuscript received June 30, 2025.}
\thanks{(Corresponding author: Wei~Song, He Zhang)}
}

% \thanks{
% % This paper was produced by the IEEE Publication Technology Group. They are in Piscataway, NJ.
% }% <-this % stops a space
% \thanks{
% % Manuscript received April 19, 2021; revised August 16, 2021.
% }

% }

\markboth{Journal of \LaTeX\ Class Files,~Vol.~14, No.~8, June~2025}%
{Shell \MakeLowercase{\textit{et al.}}: A Sample Article Using IEEEtran.cls for IEEE Journals}

\IEEEpubid{0000--0000/00\$00.00~\copyright~2021 IEEE}

\IEEEtitleabstractindextext{
\begin{abstract}
Enhancing the modular structure of existing systems has attracted substantial research interest, primarily through (1) software modularization and (2) identifying design issues (\eg smells) as refactoring opportunities; however, both approaches often prove impractical to guide effective improvement.
Inspired by both aforementioned approaches, our previous study introduced a novel and practical architectural smell -- called \emph{Pairwise Modular Smell} (or \emph{PairSmell}) -- for identifying flawed architectural decisions that necessitate further examination.
\emph{PairSmell} is defined as the deviations between the actual modular relation (MR) and the `apt MR'-- an MR agreed on by multiple modularization tools (as raters).
Although \emph{PairSmell} has shown its relevance, the reliance on external modularization tools makes it a relatively obscure concept within the community, which in turn may threaten its validity to be used in inspecting software module structure. 

The objective of this study is to explain \emph{PairSmell} from the perspective of pair characteristics. To this end, we first conduct a rapid review to collect and synthesize 19 pair characteristics that have been used in the literature to represent relationships between two entities. 
The collected characteristics are then used to train machine learning models for predicting two forms of \emph{PairSmell} -- inapt separated pairs $\mathit{InSep}$ and inapt collocated pairs $\mathit{InCol}$, based on a curated dataset of over 6,135,000 pairs of entities derived from 11 open-source Java projects. 
The trained models achieve up to a 58.6\% improvement in ROC-AUC over the baselines.
The interpretation of the models reveals that the most influential features for $\mathit{InSep}$ are out-going dependencies, terms shared with others, and declared fields; while those for $\mathit{InCol}$ include semantic similarity based on tf-idf, terms shared between the pair, terms shared with others, and in-going dependencies. 
We complement the work with a series of practical examples to illustrate how the influential pair characteristics impact the occurrence of \emph{PairSmell}.
Among our findings, a high number of out-going dependencies of a separated pair may raise questions about the separation, while a low level of shared terms may not justify collocating the two entities.

\end{abstract}

}

\maketitle

\IEEEdisplaynontitleabstractindextext

\IEEEpeerreviewmaketitle

\section{Introduction}

\emph{Software Modularity} is an essential quality attribute reflecting how a system is structured into different parts (\ie, \emph{modules}), and allowing complex software to be manageable~\cite{baldwin2000design} and reusable~\cite{kruger2020empirical}. This attribute has demonstrated a substantial impact on software reuse~\cite{kruger2020empirical}, and has been considered in various modern design scenarios, \eg microservices-based systems~\cite{abgaz2023decomposition} and LLM-enabled systems~\cite{wang2024reposvul}. Nevertheless, determining appropriate modules can be challenging in practice. This is because modules can evolve quickly~\cite{schroder2021search}, \eg due to changing functional or non-functional requirements.
Modules that worked well in the past might not fit optimally into the current system.

Substantial research effort has focused on  providing methodological support to improve the modularity of existing systems~\cite{teymourian2022fast,pourasghar2021graph,schroder2021search,candela2016using}, primarily on two directions.
On the one hand, \emph{software modularization} techniques search for a (near-) optimal modular solution to replace the original modules.
However, since such solutions often require extensive and costly changes to the original systems, developers rarely implement them.
On the other hand, some approaches identify issues (\eg anti-patterns and smells~\cite{xiao2021detecting,mumtaz2021systematic,griffith2014design}) in modular structures as refactoring opportunities in subsequent development to improve the degraded modules.
The problem is that most of the issues are coarse-grained, i.e. at the module level, making it difficult to determine refactoring strategies~\cite{CAI2023107322}.
A typical example is Cycle Dependency~\cite{fontana2017arcan}, where the chain of relations among several modules breaks the desirable acyclic nature of modules' dependency structure.
Although we know that cycle dependencies should be broken, it is difficult to decide which dependencies to break~\cite{oyetoyan2015decision}.

Building on both aforementioned directions, a novel architectural smell, \emph{Pairwise Modular Smell (PairSmell)}, was introduced in our previous study~\cite{zhong2025pairsmell} to inspect and improve the modular structure of existing software. This smell focuses on %the modular design with a specific granularity termed 
\emph{Modular Relation (MR)}, \ie whether an entity (file) pair is collocated or separated within the same module. The idea is that, if multiple modularization tools consensually designate the MR of a pair as either collocated or separated, this is considered the ground truth and the relation is termed `apt MR'. In contrast, \textit{if the actual MR of a pair violates the apt MR, this violation indicates an inappropriate architectural decision}~\cite{garcia2009identifying} --- an instance of \emph{PairSmell}.

\begin{figure}
  \centering
  \includegraphics[width=.8\linewidth]{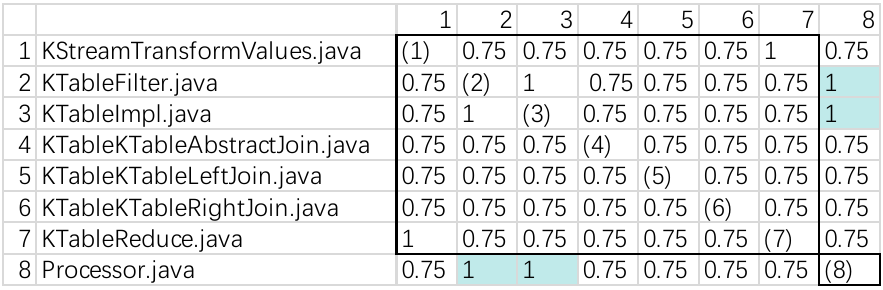}
  \caption{An InSep example in the project \emph{Kafka}. Each number indicates the average frequency that two entities are grouped together by tools. Entities in a lined rectangle actually belong to one module.}
  \label{fig: case_insep}
\end{figure}

More specifically, detailed deviations of the actual MR from the apt MR give rise to two specific forms of \emph{PairSmell}: \emph{InSep} where the apt MR for a separated pair is instead collocated, and \emph{InCol}, where the apt MR for a collocated pair is instead separated.
Fig.~\ref{fig: case_insep} depicts an instance of $\mathit{InSep}$, where \emph{Processor} is located in a separate module from all other files. However, we can see from the cells annotated with $1$ that all tools assigned it to be collocated with files \texttt{KTableFilter.java} (row 2) and \texttt{KTableImpl.java} (row 3), suggesting that the two entities might be highly related.
Our empirical study on 260,003 instances found that \emph{PairSmell} (1) is prevalent in real software projects, (2) can be detrimental by inducing 190\% more cross-module co-changes and 35\% less within-module co-changes than appropriately collocated or separated pairs respectively, and (3) can persist long if left unaddressed.
In a nutshell, \emph{PairSmell} offers fundamental insights that help developers to inspect and improve software modules more effectively.

This study enhances the potential of \emph{PairSmell}, by making it more \textbf{explainable}.
Specifically, \emph{PairSmell} is defined as the deviations between the actual MR and the `apt MR', which itself is derived using multiple modularization tools in a largely opaque or ``black box'' manner.
The output of modularization tools is used directly as the input to identify \emph{PairSmells}, following the consensus clustering strategy~\cite{fred2005combining,zhang2022weighted}.
As a result, for a pair detected as a \emph{PairSmell} instance, say \texttt{Processor.java} and \texttt{KTableFilter.java} in Fig.~\ref{fig: case_insep}, what we know is that all modularization tools agreed on their MR as collocated. However, it is hard to understand \textit{why this particular pair exhibits a PairSmell while others not} -- that is, \textit{what characteristics make some pairs more prone to being affected by PairSmell}.
Answering this question would not only deepen our understanding of \emph{PairSmell}, but also help explain why such smells are detrimental to a system's maintenance and evolution -- thereby motivating potential refactoring efforts.

The objective of this study is %to demystify \emph{PairSmell} by systematically investigating this smell. 
particularly to identify the characteristics that make certain pairs more susceptible to \emph{PairSmell}.
To achieve this, we first collect and synthesize a set of pair characteristics that can theoretically represent the relationships between two entities. 
We conduct a rapid review of 178 primary studies, from which we extract 19 distinct pair characteristics (features).
These features capture six dimensions, including \emph{relatedness}, \emph{distance}, \emph{independence}, \emph{complexity}, \emph{cohesion}, and \emph{size}. 
Next, we employ these features as independent variables, together with two metrics indicating the presence of \emph{PairSmell} -- $\mathit{InSep}$ and $\mathit{InCol}$ -- as dependent variables to train machine learning models. 

Our experiments, conducted on 6,028,271 separated and 107,606 collocated pairs, demonstrate that the selected characteristics can effectively distinguish $\mathit{InSep}$ from other separated pairs, and $\mathit{InCol}$ from other collocated pairs, with ROC-AUC values exceeding the baselines by up to 58.6\%. 
Through model interpretation, we find the most influential indicators for $\mathit{InSep}$ as 
out-going dependencies, terms shared with others, and declared fields of a pair.
For $\mathit{InCol}$, the most important indicators include semantic similarity based on tf-idf, terms shared between the pair, terms shared with others, and in-going dependencies.
In addition, we complement the quantitative analysis with a qualitative examination of representative \emph{PairSmell} examples, further illustrating the manifestations of the smell.

Overall, this study distinguishes itself from prior work by making two significant contributions.
First, this study offers a systematic characterization and explanation of \emph{PairSmell}.
% identifying the pair characteristics that help explain why specific modular relations are likely to be inapt.
While our previous work~\cite{zhong2025pairsmell} introduced and validated the concept of \emph{PairSmell}, the present study goes further by identifying and interpreting the underlying pair characteristics that explain its emergence.
By examining 19 pair characteristics, the study substantially advances the understanding of \emph{PairSmell}, offering empirical evidence that supports a more nuanced perspective beyond earlier findings, which primarily focused on its prevalence and consequences.
% clarifying why particular collocation (or separation) decisions may be questionable. This study offers empirical evidence that supports a more nuanced perspective beyond earlier findings, which primarily focused on its prevalence and consequences.
Second, the contributions of this study extend beyond \emph{PairSmell} to illuminate the relationships between pair characteristics and modular design.
% by illuminating how pair characteristics can inform modular design assessment and support future research and practice.
Through a rapid literature review to identify relevant pair characteristics, combined with predictive modeling and illustrative examples, this study systematically reveals which types of pair characteristics are associated with inappropriate—and appropriate—modular relationships. 
Consequently, the findings provide valuable insights for the modular design community, not only for managing \emph{PairSmell}, but also for leveraging pair characteristics to inform future research and practice in modular design.
% as actionable cues in architectural analysis, informing future research and practice in modular design.

% Overall, this study offers a systematic characterization and explanation of \emph{PairSmell} by examining it through the lens of pair characteristics. 
% Drawing on a rapid literature review, predictive modeling, and illustrative examples, it advances a nuanced perspective of the smell. 
% In doing so, the study makes a significant contribution to understanding \emph{PairSmell} by providing empirical evidence that substantiates its validity -- extending beyond prior insights primarily on its prevalence and consequences.
% Moreover, the findings of this study can be promisingly harnessed in the management of \emph{PairSmell}, particularly by facilitating a refactoring strategy that can be driven by pair characteristics. 

% is both characteristic-driven and instance-aware.
% including \textit{early and continuous identification}, \textit{granular, intermittent, and selective refactoring}, and \textit{whole-process modular training}.

The structure of this article is as follows. 
Section~\ref{sec: background} introduces existing techniques for improving the modularity of software systems, including \emph{PairSmell}.
Section~\ref{sec: pair characteristics to study} presents a rapid literature review to collect candidate pair characteristics for explaining \emph{PairSmell}.
Section~\ref{sec: results} elaborates the research methods and results of predictive modeling for characterizing \emph{PairSmell}. 
In section~\ref{sec: illustrative examples} we use several representative examples to illustrate the important characteristics found in Section~\ref{sec: results}.
How our findings would contribute the understanding of \emph{PairSmell} and threats to validity are discussed in section~\ref{sec: discussion} and \ref{sec: ttv}.
Section~\ref{sec:conclusions} concludes this article.

\section{Background and Related Work}
\label{sec: background}

Many studies have explored improving the modularity of existing systems, including software modularization techniques, modularity issue identification, and our \emph{PairSmell} built on both types of aforementioned methods.

\subsection{Software Modularization Techniques}

Numerous modularization techniques have been developed to restructure a large software system into smaller and more manageable subsystems~\cite{sarhan2020software}.
These techniques typically conceptualize modularization as an optimization problem, seeking an optimal solution to refactor the original modules.
The most commonly used optimization objectives are intraconnectivity (\emph{high cohesion}) and interconnectivity (\emph{low coupling}), \eg~in \cite{teymourian2022fast,yang2022enhancing,pourasghar2021graph,mitchell2006automatic}.
For instance, FCA~\cite{teymourian2022fast} is a clustering algorithm that exploits a series of operations on the dependency matrix of a system to maximize intra-dependencies within the clusters and minimize inter-dependencies between the clusters.
Another example is from the microservices domain, where Cromlech~\cite{quattrocchi2024cromlech} is presented to decompose operation and data entities into services that optimize cohesion and decoupling between services.
Additionally, some researchers incorporate refactoring effort, such as the number of changes~\cite{schroder2021search}, as an objective to minimize the effort required for modularization.
However, an industrial case study~\cite{schroder2021search} reveals that completely modularizing an entire system remains prohibitively expensive and thus impractical, given the extensive size of the code base.
\textit{Instead of seeking to restructure an entire system, the idea of PairSmell is to integrate the intelligence of multiple modularization techniques to deduce promisingly appropriate MR designs and accordingly identify inapt MRs (as opportunities that necessitate refactoring).}

\subsection{Modularity Issue Identification}

Identifying and analyzing modularity-related `issues' is an essential objective for many architecture analysis activities, such as architectural quality measurement~\cite{al2012precise} and architectural smell detection~\cite{mo2019architecture,liu2024prevalence}. 
Architecture metrics, including modularity and maintainability measures~\cite{mo2016decoupling}, aim to assess the extent
to which a software system is maintainable. 
In addition, numerous metrics of coupling~\cite{almugrin2016using} and cohesion~\cite{athanasopoulos2014cohesion} can be employed to identify quality issues at the module level. For instance, MCI~\cite{zhong2023measuring} can be used to measure which services in a system are overcoupled with others and thus might need to be refactored.

Architectural smells represent structural problems that negatively influence software evolution~\cite{xiao2021detecting,mo2019architecture} and can indicate refactoring opportunities in subsequent development. Since Joshua Garcia's definition~\cite{garcia2009identifying}, numerous types of architectural smells have been introduced within the community.
For example, Wong et al.~\cite{wong2011detecting} introduced \emph{modularity violation} which refers to two components that consistently change together but belong to separate modules.
Le et al.~\cite{le2018empirical} discovered \emph{co-change coupling}~\cite{le2018empirical} where changes to one component require changes in another component.
Mo et al.~\cite{Mo2015} presented \emph{cross-module dependency} to indicate two structurally independent modules that frequently change together in the revision history.
\textit{Compared to these smells, (1) PairSmell is defined at the fine-grained pair level, thus providing more actionable insights to enhance existing software modules; (2) while the above smells focus on the deviation between modular structure and historical revisions, PairSmell concerns deviation in the modular structure from the apt or ideal design decisions, offering a broader perspective than the existing smells.
}

\subsection{PairSmell: Definition, Identification, and Impacts }
\label{sec: pairsmell definition}

\emph{PairSmell} is defined as a triple regarding a pair of entities $e_i$ and $e_j$, where the actual MR violates its apt MR:
\begin{equation}\label{for: smell}
\!\mathit{PairSmell} =\, <(e_i,e_j), \mathit{MR_{act}} (e_i,e_j), \mathit{MR_{apt}} (e_i,e_j)> 
\end{equation}

The first element $(e_i,e_j)$ denotes a pair of entities in a target system, where $e_i \ne e_j $. 
An entity is a single code file, following our previous study~\cite{zhong2025pairsmell}.
Both the second and third elements, $\mathit{MR_{act} (e_i, e_j)}$ and $\mathit{MR_{apt} (e_i, e_j)}$, denote modular relations between entities $e_i$ and $e_j$. The MR of a pair in a specific design $d$ is separated or collocated, formally:
\begin{equation}\label{for: MR}
    \mathit{MR_{d}} (e_i, e_j) =\begin{cases}
    \text{\textit{0}}, & \text{if $mod_d(e_i) \neq mod_d(e_j)$}\\
    \text{\textit{1}}, & \text{if $mod_d(e_i)=mod_d(e_j)$}
    
\end{cases}
\end{equation}
where $mod_d(e_i)$ is the module to which $e_i$ belongs in design $d$.
$\mathit{MR_{act}(e_i, e_j)}$ is the actual modular relation of the pair, which could be extracted from a snapshot of the system.
Inspired by \emph{consensus clustering}~\cite{fred2005combining,zhang2022weighted}, an MR is considered apt if it is agreed upon by multiple modularization tools. 
In contrast, if modularization tools disagree, it suggests that the pair may be reasonably designed as either collocated or separated. 
Formally, an apt MR exists if: 
\begin{equation}
     % \mathit{MR_{apt}(e_i, e_j)} \Leftrightarrow 
     \mathit{MR_{d_1}(e_i, e_j)} = ...= \mathit{MR_{d_m}(e_i, e_j)}
\end{equation}
where $m$ is the number of modularization tools considered.

\begin{figure}
  \centering
  \includegraphics[width=\linewidth]{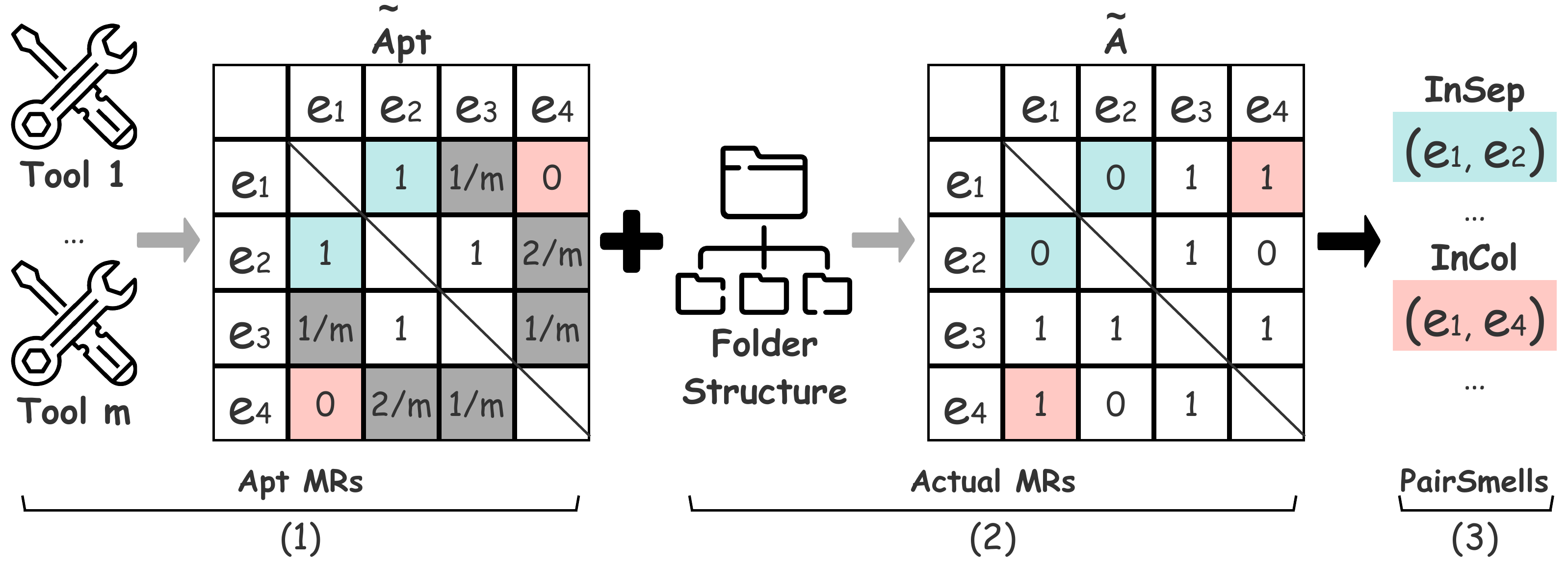}
  \caption{Overview of identifying PairSmell~\cite{zhong2025pairsmell}}
  \label{fig: overview}
\end{figure}

For \emph{PairSmell} identification, the apt MRs agreed upon by multiple modularization tools are first inferred, which are then utilized as references to identify smell instances, as illustrated in Fig.~\ref{fig: overview}. The apt MRs are inferred by comparing $m$ solutions from distinct modularization tools. The actual MRs are collected from a system's existing modules. Finally, the smell instances are detected by comparing the apt MRs with the actual MRs for each pair of entities. 

Our prior empirical study on 146,668,710 separated pairs and 3,866,940 collocated pairs from 20 C/C++ and Java projects reveal that (1) \emph{PairSmell} is prevalent among projects, with $\mathit{InSep}$ and $\mathit{InCol}$ instances covering 14.60\% and 20.44\% of the entities on average; (2) on average, entities in $\mathit{InSep}$ MRs co-change almost three times more than in other separated pairs, and entities in $\mathit{InCol}$ MRs co-change 35\% less than other collocated pairs, dramatically deviating from well-structured modules; and (3) \emph{PairSmells} persist in software projects if left unaddressed.

\section{Pair Characteristics to Study}
\label{sec: pair characteristics to study}

This section aims to collect and synthesize pair characteristics from the literature to form the candidate features for explaining \emph{PairSmell}.
In particular, our aim is to answer the following research question.

\noindent
\textit{\textbf{RQ1. What pair characteristics can be used to represent the relationship between two entities?}}

This question is to identify the characteristics of entity pairs that can be used to describe their relationships and are thus pertinent to \emph{PairSmell}; the answer can provide a solid theoretical basis for our study.
We answer this question using the rapid literature review as the high-level method for data collection, and thematic synthesis as the data analysis method for synthesizing pair characteristics.

\subsection{Methodology}

We follow the guidelines of Shull et al.~\cite{shull2008guide} and adopt scientific literature as the source of data collection, as it offers greater rigor and credibility to the collected evidence. 
Although this research studies the modular design for \textit{a pair of entities}, our initial searches (\ie with Scopus, IEEE Xplore Digital Library and ACM Digital Library) indicated that only a limited set of studies explore the modular design of software specifically at this level. Therefore, we broadened the scope of the search and analyzed the literature whose subject is the modular design for \textit{multiple entities}, \aka `software modularization techniques'.

Specifically, we use rapid literature review~\cite{tricco2015scoping} -- a form of evidence collection and synthesis that simplifies the process of a systematic literature review to quickly acquire knowledge. To retrieve the primary studies, we consulted the articles from the two most up-to-date, systematic reviews related to software modularization: Sarhan et al.~\cite{sarhan2020software} collected 143 papers that examined the clustering of software modules; and Abgaz et al.~\cite{abgaz2023decomposition} collected 35 papers that studied the decomposition of monolithic applications into microservices. 
By combining the two sets, we obtained 178 primary studies for analysis.
We retained all 178 studies without further filtering, as they have all been included in recent secondary studies and collectively represent the state of the art in software modularization. This approach was intended to preserve as much evidence as possible.

Since many primary studies discussed the relationships between entities as the basis for modularization, we extracted these relationships as pair characteristics, along with the underlying rationales. These extracted items help to investigate respectively (1) \textit{what pair characteristics can be used to determine the modular structure} and (2) \textit{why}. 
At the same time, we also extracted the core objective of software modularization that is discussed by researchers to further guide our synthesis of pair characteristics.
Note that while the core objective outlines the desired attributes of software modules (according to the researchers), the underlying rationales for each pair characteristic explain why the characteristic contributes to achieving the modularization objective.
The research methods and procedures are detailed below.

We synthesized pair characteristics from primary studies using thematic synthesis~\cite{Cruzes2011}.
According to Cruzes and Dyba~\cite{Cruzes2011}, general thematic synthesis steps include: extracting data ($S_1$), coding data ($S_2$), translating codes into themes ($S_3$), creating a model of higher-order themes ($S_4$), and assessing the synthesis' trustworthiness ($S_5$). 
Our thematic analysis process was inductive, as there is no predefined taxonomy organizing current pair characteristics.

In $S_1$, the primary studies were assigned to two researchers (the 1st and 4th authors) with prior experience in analyzing qualitative data. The goal was to extract pair characteristics, rationales, and objectives by reading each study's abstract, introduction, and other sections (\eg the approach section) if required. 
Two researchers started by extracting data from 20 studies together to coordinate on the task and align their understanding of the extracted items (\eg pair characteristics).
After that, each researcher independently extracted data from all the remaining studies and subsequently merged the results with any disagreements resolved through discussion with the other. We did not calculate a consistency score (\eg Kappa~\cite{kitchenham2007guidelines}) as the extracted data is mostly qualitative without predefined categories.

In $S_2$, two researchers independently performed \emph{open coding} for all papers, assigning a code whenever a concept became apparent. For example, one primary study~\cite{glorie2009splitting} stated: \textit{``In order to make sure that highly related building blocks appear in the same concept ... We consider a building block A to be dependent on a building block B if A uses a function or data structure in B''}. 
When coding this data, the researchers used the code \textit{`method invocation'} \textit{(for ``A uses a function in B''}) and the code \textit{`data accessing'} (for \textit{``A uses a data structure in B''}) as two pair characteristics relevant to modular design.

In $S_3$, we first excluded the pair characteristics that cannot be directly retrieved from the source code (\eg the composition relation from UML diagrams, and co-change from commit analysis).
Our focus is on analyzing modular design based on the static code structure, as it directly reflects the modules in the development environment.
We then performed \emph{axial coding} to translate codes into themes by comparing and merging similar items.
For example, \textit{`method invocation'} and \textit{`data accessing'} were translated into \textit{`structural dependency'} (\textit{`dependency'} in Table~\ref{tab: characteristics}).

In $S_4$, \emph{selective coding} was performed to output the core themes (\ie pair characteristics) by organizing the themes into a hierarchy structure. 
We reviewed the themes and interpreted their relationships, placing more important themes at higher levels of the hierarchy.
For example, structural dependencies, common terms, and similarities between two entities were grouped under `relatedness', as they collectively describe the closeness between entities.
In $S_5$, we validated the core themes by assessing their trustworthiness against the original data.
Specifically, we examined the data to determine how well the themes captured it and how their relationships aligned with the data.

\subsection{Results}

The final results can be found in our online repository~\footnote{\url{https://figshare.com/s/326e080186e95a70f859}\label{foot: online data}}, which covers pair characteristics relevant to six core themes derived from the thematic synthesis: \emph{relatedness}, \emph{distance}, \emph{independence}, \emph{complexity}, \emph{cohesion}, and \emph{size}. 
The theme \emph{relatedness} consists of the pair characteristics that describe the connection (\eg structural dependencies and semantic similarity) between two entities.
\emph{Distance} describes the degree of difference between two entities (\eg if they are linked by a dependency). 
\emph{Independence} emphasizes to what extent two entities are independent in the system.
Finally, \emph{cohesion}, \emph{complexity}, and \emph{size} denote the average cohesion, complexity, and size of two entities.

Table~\ref{tab: characteristics} shows a list of 19 selected characteristics that cover the six core themes, together with the rationale for each characteristic in the context of modular design. 
We highlight that compared to the review results (see the appendix), we made three additional changes: (1) Although there are various types of dependencies between two entities (\eg method invocation and data accessing), we merged these different dependencies together to provide a more comprehensive view of structural dependency. 
(2) While the review identified complexity, cohesion, and size as relevant factors for modular design, the studies did not specify detailed metrics; consequently, we adopted commonly used metrics related to these attributes to describe the corresponding characteristics of a pair. 
(3) We deliberately adjust certain characteristics to ensure their suitability, and to maintain consistency between different themes. For instance, we excluded `same folder' because the modular relation (collocated or separated) already captures whether two entities belong to the same folder.
Please see our online repository~\ref{foot: online data} for details.

While we conducted a thematic synthesis to organize the themes, 
some overlap between characteristics across themes may still exist. For instance, structural dependencies are utilized to measure \emph{dependency} (within the Relatedness theme), \emph{str-distance} (within the Distance theme), and \emph{in-degree} (within the Independence theme).
This overlap is partially due to our inductive approach~\cite{Cruzes2011}, in which we assigned themes based on concepts that naturally emerged from the data.
For example, many primary studies emphasize the importance of relatedness between entities — such as the number of structural dependencies — as a crucial factor in determining appropriate software modules. To mitigate the potential impact of feature overlap on predictive modeling, we performed a collinearity analysis to remove collinear features, as detailed in Section~\ref{sec: model fitting}.

\begin{table}[tbh!]
\caption{Selected Pair Characteristics to Study}
\label{tab: characteristics}
\scriptsize
\centering
\begin{tabular}{@{}lp{240pt}@{}}
\toprule
% \toprule
\textbf{T.} & \textbf{Description and Rationale}   \\ 
\midrule
% \midrule
% \multicolumn{3}{c}{\textbf{At Pair Level}}\\
% \midrule
\multirow{6}{*}{\rotatebox[origin=c]{90}{Relatedness}}  & \textbf{dependency}: The number of structural dependencies between two entities. \\
  % & \underline{Rationale}: The more structural dependencies between two entities, the more likely that they should be collocated.\\
  % \cmidrule{2-2}
 & \textbf{intersection}: The number of common terms in the code files of two entities.\\
 % & \underline{Rationale}: Two entities with more common terms are more likely collocated. \\
 % \cmidrule{2-2}
& \textbf{sim-tfidf}, \textbf{sim-tm}: The similarity between two entities calculated using tf-idf and topic modeling. \\
& \underline{Rationale}: Two entities that are more related (according to their dependencies and similarity) are likely to be collocated~\cite{wang2017automatic}.\\
% & \underline{Rationale}: Two similar entities are likely to be collocated.\\
\midrule
\multirow{5}{*}{\rotatebox[origin=c]{90}{Distance}}  & \textbf{str-distance}: The distance between two entities in a pair. The value can be 1 if there is a direct dependency between them; otherwise, INF. \\
 & \textbf{sem-distance}: The distance between two entities according to their common terms. The value can be 1 if they have common terms; otherwise, INF. \\
& \underline{Rationale}: Two entities that are close to each other are more likely to be collocated than others that are distant.
\\
\midrule

\multirow{6}{*}{\rotatebox[origin=c]{90}{Independence}}  & \textbf{in-degree, out-degree}: The average number of dependencies going into/out the two entities in a pair. \\
 & \textbf{degree}: The average number of common terms the two entities share with all other entities in the system.\\
 & \underline{Rationale}: The average number of dependencies of two entities and the average number of common terms can reflect their independence in the system, which may probably affect their modular relation. \\

\midrule
\multirow{14}{*}{\rotatebox[origin=c]{90}{Complexity}} &  \textbf{cbo}: Coupling between objects. The average number of dependencies coupled with each entity of a pair. \\
& \textbf{wmc}: Weight Method Class. The average complexity of each entity, calculated as the sum of the McCabe’s cyclomatic complexity of its methods.\\
& \textbf{dit}: Depth Inheritance Tree. This metric counts the average number of fathers each entity has. All classes have DIT at least 1.\\
& \textbf{rfc}: Response for a class. This metric counts the average number of unique method invocations in each entity.\\
& \textbf{max-nested-block}: Highest number of code blocks nested together in a file.\\
& \textbf{comment-density}: The average percentage of lines in an entity containing either comment or commented-out code.\\
 & \underline{Rationale}: Empirical studies show that poorly modularized artifacts correlate with higher complexity. \\
 \midrule
\multirow{5}{*}{\rotatebox[origin=c]{90}{Cohesion}} & \textbf{lcom}: Lack of cohesion in methods. This metric counts the average number of methods in each entity that are not related (through the sharing of some of the entity's fields).\\
& \underline{Rationale}: A highly cohesive entity is internally focused and self-contained, thus it is less likely to benefit from being clustered with other entities. \\
% The cohesion of the entities may have an impact on the likelihood of the pair to be collocated. \\
\midrule
\multirow{5}{*}{\rotatebox[origin=c]{90}{Size}} & \textbf{total-methods}, \textbf{total-fields}: Total number of methods and declared fields in a file.\\
& \textbf{ncloc}: The average number of lines of code in a file, ignoring empty lines. \\
& \underline{Rationale}: The size of the entities may have an impact on the likelihood of the pair to be collocated. \\
% \bottomrule
\bottomrule
\end{tabular}
\begin{tablenotes}
        \item * ``T.'' Stands for Themes.
\end{tablenotes}
\end{table}

All the 19 pair characteristics can be collected automatically by leveraging tools for syntactic and semantic analysis.
Specifically, to collect these pair characteristics, \emph{Depends}~\cite{depends2022} was used to recover structural dependencies, as it is capable of extracting 13 dependency types by analyzing the syntactic structures of the source code, such as \emph{call}, \emph{contain}, and \emph{implement}. 
The structural dependencies also serve as the basis for collecting other characteristics, including \emph{in-degree} and \emph{out-degree}.
For collecting semantic-related features, the \emph{Ctags}~\cite{ctags2022} tool was used to extract identifiers (\eg function and variable names) found in source code files, which were then analyzed with natural language processing techniques including topic modeling (via the NLTK~\cite{nltk2023} toolkit) and TF-IDF (via the scikit-learn~\cite{scikit2024} library).
In addition, a static analysis tool called \emph{CK}~\cite{ck2022} was leveraged to calculate the complexity, cohesion, and size metrics of each entity (\eg \emph{dbo}, \emph{dit}, and \emph{lcom}).
Finally, an open-source tool \emph{cloc}~\cite{cloc2025} was used to count comment lines (\emph{comment-density}) and source code lines (\emph{ncloc}) for each entity. The measurements at entity level were then aggregated at pair level using the \emph{average} function to derive the corresponding characteristics of a pair.

\section{Pair Characteristics That Matter}
\label{sec: results}

In this section, our objective is to characterize and explain \emph{PairSmells}: using the pair characteristics identified in Section~\ref{sec: pair characteristics to study}, we aim to determine if the characteristics contribute to the occurrence of \emph{PairSmell}, and if so, which ones contribute more.
We use the following questions to guide the exploration.

\noindent
\textit{\textbf{RQ2. Can the pair characteristics be used as indicators for \emph{PairSmell}?}}

\noindent
\textbf{\textit{RQ3. Which pair characteristics are the most important indicators for \emph{PairSmell}?}}

Specifically, RQ2 aims to explore the association between pair characteristics and the occurrence of \emph{PairSmell}.
To answer this question, we train several machine learning models using features collected and derived from pair characteristics. A positive answer to this question is the prerequisite for exploring RQ3 which aims to identify the pair characteristics most associated with \emph{PairSmell}.
To answer RQ3, we identify the most important characteristics of the prediction model and analyze how a change in these characteristics affects the model’s prediction.
Such an exploration can help us to better understand the type of pairs that fall into the \emph{PairSmell} group.

Considering the selected pair characteristics as independent variables, this section first describes the other side of machine learning models, \ie the design of dependent variables. After that, it elaborates on our procedures for curating datasets, fitting models to characterize \emph{PairSmell}, and analyzing the results to answer each research question.

\subsection{Dependent Variables}

We selected two metrics to serve as representations of \emph{PairSmell} according to its definition (cf. Section~\ref{sec: pairsmell definition}); the metrics were identified in our previous work~\cite{zhong2025pairsmell}. %Each metric indicates whether a pair is affected by a specific form of \emph{PairSmell}. 

\textbf{Inapt Separated (InSep)}---two entities are separated into different modules in the actual system, but the \emph{apt} MR is collocated according to modularization tools. This smell means that the two separated entities are highly related despite being separated, \eg due to interdependency. The inapt MR of these two entities may hamper the independence of the corresponding modules, making changes of one module propagating to another module~\cite{arvanitou2015introducing}.

\textbf{Inapt Collocated (InCol)}---two entities are actually implemented as collocated, but the \emph{apt} MR by all tools is to separate them. This smell indicates that the two entities, while placed in the same module, are not justifiably related, \eg they address different (or even orthogonal) concerns. This inapt MR may impede the cohesion of the current module, violating the single responsibility principle~\cite{martin2018clean}.

According to the above definitions, a separated pair may suffer from $\mathit{InSep}$, while a collocated pair may suffer from $\mathit{InCol}$.
To represent each of these forms as a \emph{PairSmell} metric to be predicted, we create a binary metric for each form: $p_{sep}$ for a separated pair, and $p_{col}$ for a collocated pair. 
If a separated pair is affected by $\mathit{InSep}$, then its metric $p_{sep}$ = 1; otherwise, $p_{sep}$ = 0. Similarly, if a collocated pair is an $\mathit{InCol}$, then the metric $p_{col}$ = 1; otherwise, $p_{col}$ = 0.

\subsection{Data Collection and Sampling}

We begin with the dataset collected in our previous study~\cite{zhong2025pairsmell} that explicitly labels instances of $\mathit{InSep}$ and $\mathit{InCol}$. 
Within the dataset, 20 projects were selected as they are non-trivial with at least 100 entities (with an average of 1,010.5 and 1,456 maximum), so that studying modularity smells in these projects is meaningful~\cite{liu2024prevalence}. 
In this study, we limit our analysis to 11 java projects among the 20 ones, since one used tool (\emph{CK}~\cite{ck2022}) is restricted to this programming language. The final 11 selected projects are shown in Table~\ref{tab: systems}, together with their number of entities (\#Entity), structural dependencies between the entities (\#Link), and commits (\#Cmt). These projects differ in scale, evolution histories, and business domains.

Our initial dataset consists of 6,028,271 separated pairs with 3,283 $\mathit{InSep}$ instances and 107,606 collocated pairs with 1,838 $\mathit{InCol}$ instances. Note that the number of separated pairs is much higher than that of collocated pairs due to combinatorial growth, as large systems are often intentionally decomposed into many modules to manage complexity~\cite{newman2021building}.
In addition, our dataset is imbalanced in terms of classes, \eg the rate of $\mathit{InSep}$ and other separated pairs is 1:1,835 -- which is a common phenomenon in software engineering research. 
Given that the minority class (\ie \emph{PairSmell}) is of greater interest than the majority -- we are more concerned with inapt modular relations than with apt ones (to improve the modular structure) -- we decided to conduct class balancing.
This approach allows the machine learning models to focus on the minority class, enabling them to learn features that are more relevant to \emph{PairSmell}.

In addition, we observed that some entities may occur in the dataset more frequently than other entities. For instance, a single entity in a module can be separated with all other entities in the system and thus appear $n-1$ times in the dataset. As a result, if we use the entire population as the dataset, the learned models could be biased towards these frequent entities, which might lead to overfitting on these frequent entities and underfitting on those that appear less frequently. Therefore, we adopted Inverse Transform Sampling~\cite{devroye1986sample} to generate samples from the entire population to ensure that 1) less frequent entities have a higher chance of being selected compared to more frequent ones and 2) pairs from the minority class have a higher chance of being selected compared to that from the majority class. For determining the sampling size, we use 99\% confidence level and 1\% error margin.
We further assessed the robustness of our main findings by varying the sampling strategy and found that the significant pair characteristics remain consistent. Detailed results are reported in the online repository~\ref{foot: online data}.

We acknowledge that the sampling process may affect our analysis, as different samples can lead to variations in the results. To account for this, we repeated the sampling process three times and conducted experiments using each of the resulting samples. 
For clarity, we present only the results from the first experiment in the main text, while visualizations of the results from the other two experiments are provided in the appendix~\ref{foot: online data}. Nonetheless, our overall conclusions are drawn based on all three experiments.

Our final dataset (for the first experiment) consists of 16,541 separated pairs and 14,372 collocated pairs spanning 11 projects. Among these separated pairs, 3,109 are $\mathit{InSep}$ instances, and among the collocated pairs, 1,387 are $\mathit{InCol}$ instances.
Readers may refer to the online data to view the datasets from the other two experiments.

\subsection{Model Fitting}
\label{sec: model fitting}

Our objective is to study what characterizes \emph{PairSmell}. Thus, we train machine learning models to explore the association between pair characteristics and \emph{PairSmells}. 

Collinear features can distort each other's importance in the model~\cite{mcintosh2018fix}. Therefore, we first analyze collinearity among the selected characteristics (features) using Spearman's $\rho$ rank correlation~\cite{spearman1961proof}. We chose a rank correlation because it enables to detect nonlinear correlations. Similarly to prior studies~\cite{tan2020first, weeraddana2024characterizing}, we use $\rho = 0.7$ as the threshold to remove collinear features. That is, any pair of features with $\rho \geq 0.7$ should have one of the features removed prior to the interpretation of the model.

The correlations among the pair characteristics can be found in our online repository~\ref{foot: online data}. According to the correlations, we removed the following seven characteristics when fitting the models for separated pairs: \emph{intersection}, \emph{sem-distance}, \emph{cbo}, \emph{wmc}, \emph{rfc}, \emph{lcom}, and \emph{ncloc}.
We also removed \emph{str-distance}, \emph{cbo}, \emph{wmc}, \emph{rfc}, and \emph{total-methods} when fitting the models for collocated pairs. 
When removing features from a feature set, we do so in a manner that retains as many features as possible. For example, because \emph{rfc} is highly correlated ($\geq0.7$) with both \emph{out-degree} and \emph{max-nested-blocks}, we remove \emph{rfc} and retain the other two features.
The details on reasons for choosing one feature over the other can be seen in the repository.
We observe that the pairs of features with $\rho \geq 0.7$ are consistent among the three experiments, demonstrating small variations between the three samples.

\begin{table}
\caption{Summary of the Studied Software Projects}
\label{tab: systems}

\scriptsize
\begin{tabular}{p{6pt}p{25pt}p{72pt}p{18pt}rrr}
\toprule
\textbf{$P_i$} & \textbf{Project} & \textbf{Domain} & \textbf{Version} & \textbf{\#Entity} & \textbf{\#Link} & \textbf{\#Cmt} \\
\midrule
$P_1$ &Cassandra & Row store & 0.6.10 & 283 & 5,569 & 1,752 \\
$P_2$ &Druid & Analytics database & 0.7.0 & 1,045 & 7,651 & 4,980 \\
$P_3$ &Gobblin & Data management & 0.9.0 & 1,279 & 9,743 & 3,717 \\
$P_4$ &Hadoop & Distributed framework & 0.20.0 & 890 & 17,266 & 3,461 \\
$P_5$ &Hbase & Storage system & 1.0.2 & 1,456 & 34,968 & 10,061 \\
$P_{6}$ &Iotdb & Data management & 0.11.0 & 836 & 19,273 & 4,209 \\
$P_{7}$ &Kafka & Event streaming & 0.10.2.1 & 747 & 11,593 & 3,247 \\
$P_{8}$ &Lucene & Search engine & 2.9.2 & 1,006 & 21,377 & 4,042 \\
$P_{9}$ &Mahout & DSL framework & 0.6 & 1,052 & 12,939 & 2,269 \\
$P_{10}$ &Ozone & Object store & 1.0.0 & 1,380 & 8,595 & 2,698 \\
$P_{11}$ &Pulsar & Pub-sub messaging & 2.3.0 & 1,142 & 20,519 & 2,892 \\
\bottomrule
\end{tabular}

\end{table}

\begin{table}
\caption{The Initial Dataset Collected From All 11 Projects}
\label{tab: smell proportion}
\scriptsize
\centering
\begin{tabular}{p{6pt}rr r@{\hspace{0.001cm}} rr}
\toprule
 & \multicolumn{2}{c}{\textbf{Separated Pairs}} &  &\multicolumn{2}{c}{\textbf{Collocated Pairs}} \\
 \cmidrule{2-3} \cmidrule{5-6}
\multirow{-2}{*}{\textbf{$P_i$}} & \textbf{\#InSep} & \textbf{\#All} &  & \textbf{\#InCol} & \textbf{\#All}  \\
\midrule
{$P_1$} & 27	& 37,704   && 365 & 2,199  \\
{$P_2$} & 85	& 536,552  & &90	& 8,938	  \\
{$P_3$} & 80	& 810,917  && 61	& 6,364	  \\
{$P_4$} & 334 &	380,567	  && 162	& 15,038 \\
{$P_5$} & 960 &	1,026,478	  && 176	& 22,097  \\
{$P_{6}$} & 134 &	345,827  && 189	&3,203\\
{$P_{7}$} & 93&	268,954  & &66&	9,677	 \\
{$P_{8}$} & 398	& 485,625 & &536	& 19,890  \\
{$P_{9}$} & 987&	547,890  && 36	&4,936	  \\
{$P_{10}$} & 119	& 944,732 && 44	&6,778 \\
{$P_{11}$} & 66	& 643,025  && 113	&8,486	 \\
\midrule
{\textbf{Avg.}} & 298	& 548,025 && 167	& 9,782	\\
\bottomrule
\end{tabular}
\end{table}

\subsection{Can the pair characteristics be used as indicators for \emph{PairSmell}? (RQ2)}

\textbf{Approach.} To study the relevance of pair characteristics to the occurrence of \emph{PairSmell}, we use the characteristics in Table~\ref{tab: characteristics} to train several machine learning models. 
The trained models aim to predict the appearance of \emph{PairSmell} for each pair. 
Specifically, two types of binary classifiers are trained: one to determine whether a separated pair is $\mathit{InSep}$, and another to determine whether a collocated pair is $\mathit{InCol}$.

We evaluate the performance of the trained models based on (a) the discriminatory power and (b) the ability to balance precision and recall.
The discriminatory power of the models is estimated using the \emph{Receiver Operating Characteristics-Area Under the Curve (ROC-AUC)}~\cite{hanley1982meaning} metric. The ROC is a probability curve and ROC-AUC is a value between 0 and 1 that represents the degree of which the model is capable of distinguishing between classes. The higher the ROC-AUC, the better the model is at correctly predicting classes. 
The \emph{Area Under Precision-Recall Curve (AUPRC)}~\cite{boyd2013area} is calculated to measure the models' ability to balance precision and recall across different probability thresholds. The AUPRC is a value between 0 and 1. The higher the AUPRC, the better the model is at balancing precision and recall.

We conducted preliminary experiments with \emph{Logistic Regression (LR)}~\cite{hosmer2013applied}, \emph{Support Vector Machines (SVM)}~\cite{awad2008support}, and \emph{Random Forest (RF)}~\cite{biau2016random} to compare their performance using ROC-AUC and AUPRC. 
We selected these models because our study is to understand the association between pair characteristics and the occurrence of \emph{PairSmell}, which requires models with strong descriptive capabilities that provide insight into the data.
In addition, good performance of the trained models is required in order to derive meaningful associations. 
We performed data normalization, which is important for LR and SVM when there is high cardinal variance between the features.
All three models were trained on 80\% of our dataset (training set) and evaluated on the rest 20\% (test set). 
We tuned the hyper-parameters of each model using 10-fold cross-validation on the training set and used the best hyper-parameters across folds for training the models.
Model performance was assessed on the training set using repeated stratified 10-fold cross-validation (3 repeats). Finally, each model was retrained on the full training set and evaluated once on the held-out test set to obtain the final performance results.
As presented in Table~\ref{tab: performance for candidate models}, the SVM models demonstrate relatively strong performance across both tasks (in fact, they also achieve the best performance in the other two experiments~\ref{foot: online data}). Therefore, they were selected to explore the characteristics of \emph{PairSmell}.
%Note that SVM also achieves relatively strong performance in the other two experiments.

\begin{table}
\caption{Comparison of Performance for Candidate Models}
\label{tab: performance for candidate models}
\scriptsize
\centering
\begin{tabular}{p{36pt}rr r@{\hspace{0.001cm}} rr}
\toprule
 & \multicolumn{2}{c}{\textbf{Separated Pairs}} &  &\multicolumn{2}{c}{\textbf{Collocated Pairs}} \\
 \cmidrule{2-3} \cmidrule{5-6}
\multirow{-2}{*}{\textbf{Models}} & \textbf{ROC-AUC} & \textbf{AUPRC} &  & \textbf{ROC-AUC} & \textbf{AUPRC}  \\
\midrule
LR & 0.648	& 0.412   && 0.778 & 0.385  \\
SVM & 0.673	& 0.44  & &0.793	& 0.459	  \\
RF & 0.612	& 0.369 && 0.812	& 0.649	  \\
\midrule
SB & 0.485	& 0.262  && 0.507	& 0.166  \\
CB & 0.5	& 0  && 0.5	& 0	  \\
TB & 0.5    & 0.188 &&  0.5 & 0.097 \\
\bottomrule
\end{tabular}
\begin{tablenotes}
        \item * ``LR'', ``SVM'', and ``RF'' represent logistic regression, support vector machine, and random forest, respectively.
        \item * ``SB'', ``CB'', and ``TB'' represent stratified baseline, constant baseline, and theoretical baseline, respectively.
\end{tablenotes}
\end{table}

Since there is no previous work on the use of pair characteristics to predict \emph{PairSmell}, the results are compared with three baselines: the stratified baseline, the constant baseline, and the theoretical baseline. The stratified baseline uses the class distribution in the training set for weighted random predictions about the occurrence of \emph{PairSmell}. The constant baseline uses constant results (with value 0). Finally, the theoretical baseline represents random guess, whose value for ROC-AUC is 0.5 and the value for AUPRC is determined by the positive class prevalence~\cite{saito2015precision}, \ie $\frac{tp}{tp+tn}$.

\noindent
\textbf{Results.} Table~\ref{tab: performance for candidate models} shows the evaluation results regarding ROC-AUC and AUPRC. 
The ROC-AUC for the baseline models round up to 0.5, which is the expected ROC-AUC values when the model makes random predictions or always predicts the same class. 
Compared to baseline models, we can observe that SVM models improve the ROC-AUC by 34.6\% for separated pairs and by 58.6\% for collocated pairs.
The high ROC-AUC improvement shows that the pair characteristics in Table~\ref{tab: characteristics} can be leveraged to predict if a pair is affected by \emph{PairSmell}. In other words, they can be used as indicators of \emph{PairSmell}. 

\begin{tcolorbox}[boxsep=0pt,left=3pt,right=6pt,top=4pt,bottom=1pt] \itshape
\textbf{Finding \#1:} 
The evaluation of our classification models shows that pair characteristics can discriminate effectively between the instances of InSep and other separated pairs, as well as the instances of InCol and other collocated pairs.
\vspace*{0.5ex}
\end{tcolorbox}

Regarding the AUPRC metric, we see that SVM models yield an AUPRC of 0.44 for separated pairs and an AUPRC of 0.459 for collocated pairs, surpassing the best baselines with an improvement of 1.57 times and 2.77 times, respectively. 
Similar results can be observed from the other two experiments (see the appendix~\ref{foot: online data}). 
This evidence supports the model's effectiveness in distinguishing positive instances and minimizing false positives, which is especially crucial in our dataset of separated and collocated pairs with an imbalanced class distribution.

% \vspace*{-1.0ex}
\begin{tcolorbox}[boxsep=0pt,left=3pt,right=6pt,top=4pt,bottom=1pt] \itshape
\textbf{Finding \#2:} 
Our classification models demonstrate a commendable balance between precision and recall.
\vspace*{0.5ex}
\end{tcolorbox}

\subsection{Which pair characteristics are the most important indicators for \emph{PairSmell}? (RQ3)}
\label{sec: RQ3}

\textbf{Approach.} 
Pair characteristics (\ie features) which have a greater impact on the model's prediction of the occurrence of \emph{PairSmell} are better indicators of \emph{PairSmell}. To calculate the feature importance in our models, we use the permutation feature importance~\cite{molnar2020interpretable} instead of the default feature importance of SVM.
While the default feature importance reflects how much a model relies on each input feature for predictions, permutation importance measures the performance drop when the values of a single feature are randomly permuted.
We chose the permutation feature importance because it is model agnostic and can show the importance of a feature across models.
We calculate the 10-fold permutation importances by randomly permuting each feature 10 times and observing its impact on the models' performance (ROC-AUC score). A feature (pair characteristic) is deemed more important if permuting its values has a greater impact on the models’ performance.

\begin{figure*}[t]
    \centering
    \subfloat[The important features for predicting InSep]{\includegraphics[width=.5\textwidth]{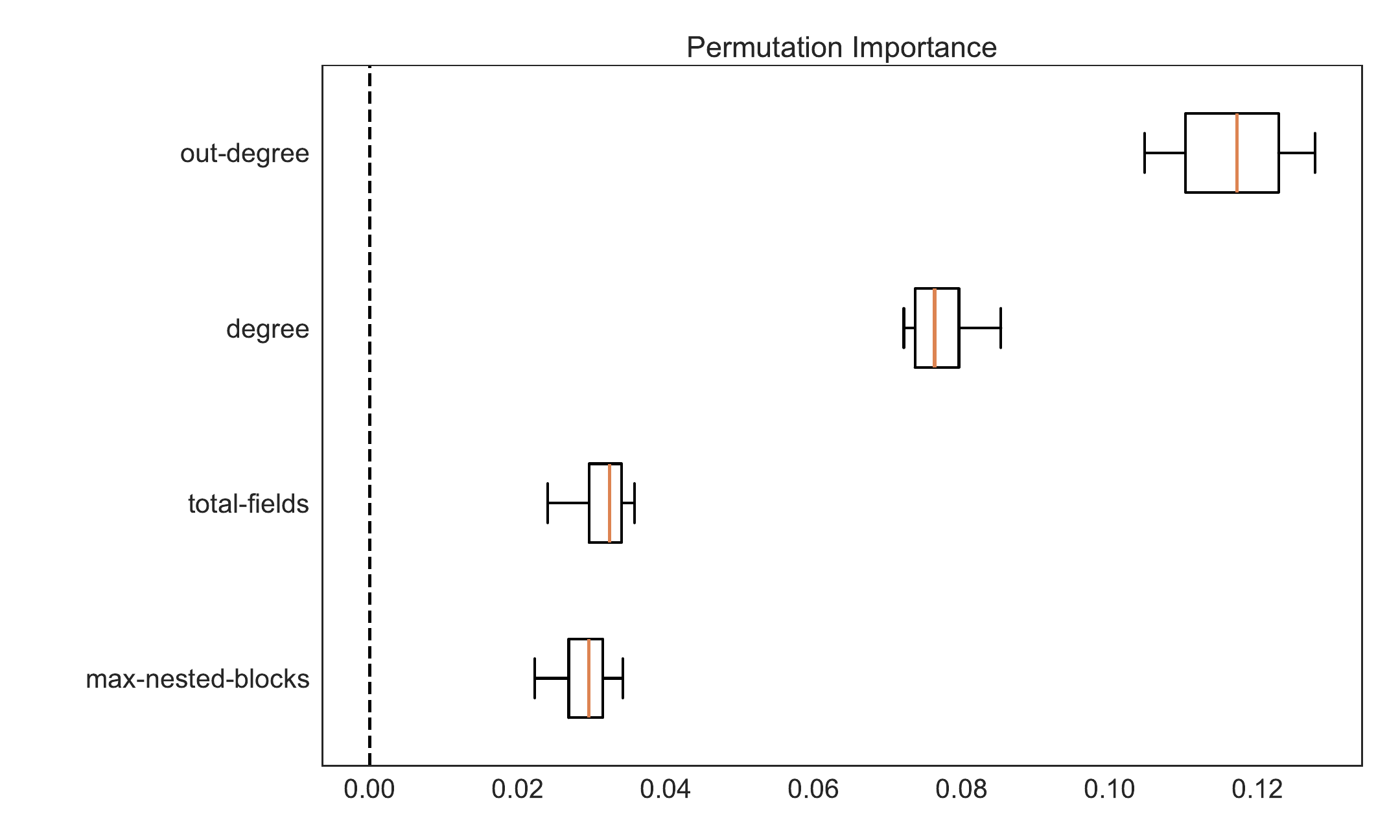}%\caption{fig1}
    }
    % \quad
    \subfloat[The important features for predicting InCol]{\includegraphics[width=.5\textwidth]{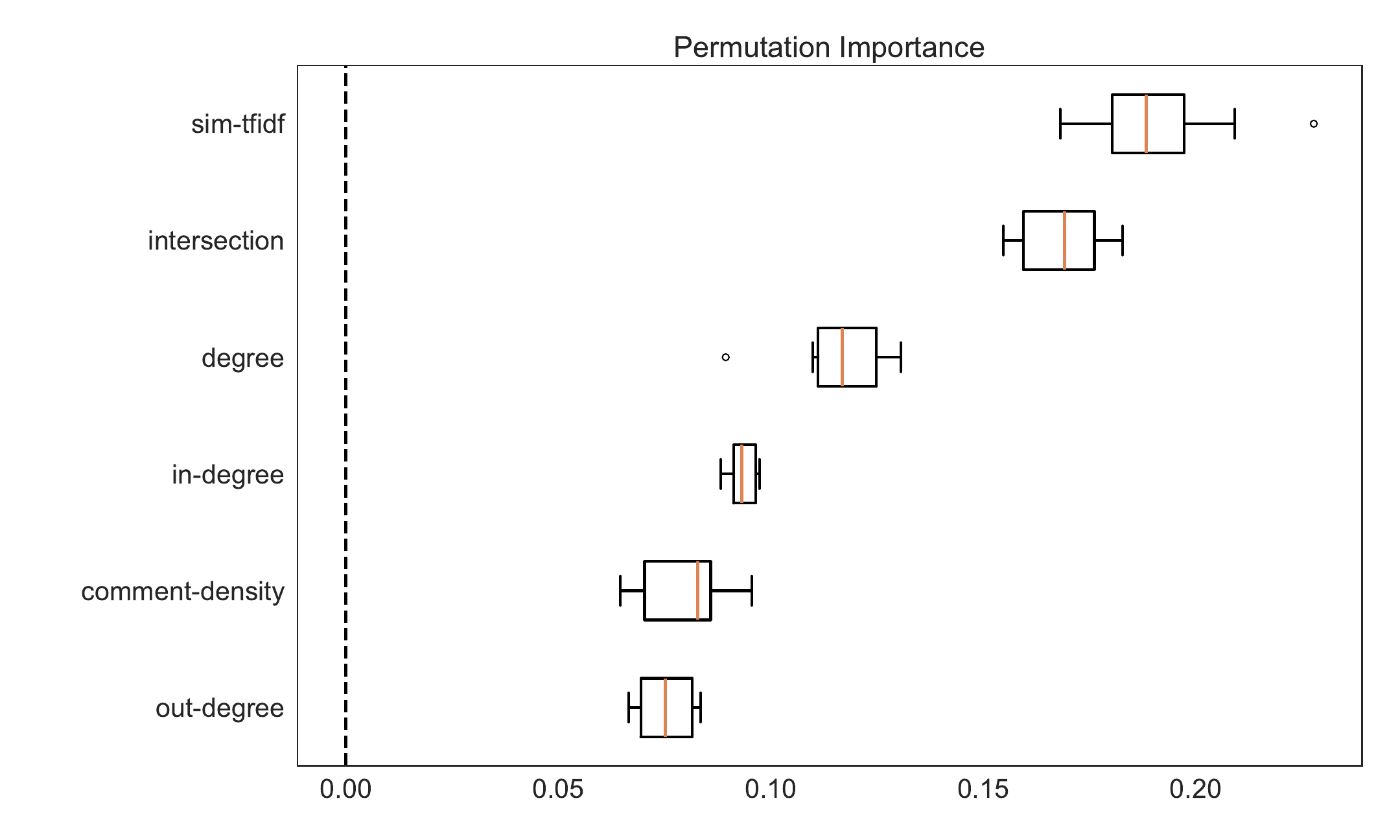}%\caption{fig1}
    }
    \quad
    \caption{The most important features associated with the occurrence of InSep and InCol} 
    \label{fig: importance features}
\end{figure*}

\noindent
\textbf{Results.} Fig.~\ref{fig: importance features} presents the features whose permutation importance is higher than zero to predict the occurrence of $\mathit{InSep}$ and $\mathit{InCol}$.
We observed that \emph{out-degree}, \emph{degree}, and \emph{total-fields} are the most important indicators of the occurrence of $\mathit{InSep}$, as they consistently rank among the top 4 indicators across all three experiments (see again appendix~\ref{foot: online data}).
This hints that these features are highly relevant to determining whether the modular design of a separated pair is appropriate. 
% The \emph{out-degree} metric is the top 1 relevant feature for the model of $\mathit{InSep}$. This suggests that the number of out-going dependencies from a separated pair may reflect how the two entities reach out to (depend on) the rest of the system, thus affecting the appropriateness of their separation (illustrated as follows).
Similarly, in terms of collocated pairs, \emph{sim-tfidf}, \emph{intersection}, \emph{degree}, and \emph{in-degree}, are the most important indicators of the occurrence of $\mathit{InCol}$ (as they rank among the top 4 indicators in all three experiments).
Please note that we chose the top 4 indicators to keep our analysis of $\mathit{InCol}$ consistent with that of $\mathit{InSep}$.
% and $\mathit{InCol}$ consistent.

\begin{tcolorbox}[boxsep=0pt,left=3pt,right=6pt,top=4pt,bottom=1pt] \itshape
\textbf{Finding \#3:} 
The most important indicators for InSep among the studied pair characteristics are out-degree, degree, and total-fields, while those for InCol are sim-tfidf, intersection, degree, and in-degree.
\vspace*{0.5ex}
\end{tcolorbox}

\begin{figure*}[t]
    \centering
    {\includegraphics[width=.24\textwidth]{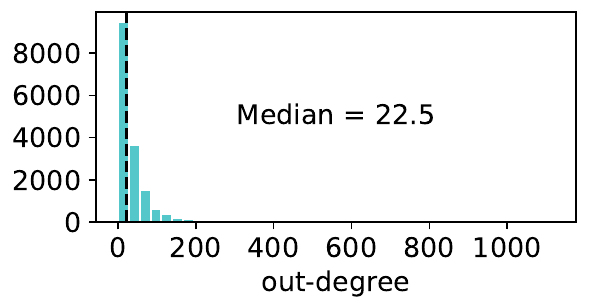}%\caption{fig1}
    }
    {\includegraphics[width=.24\textwidth]{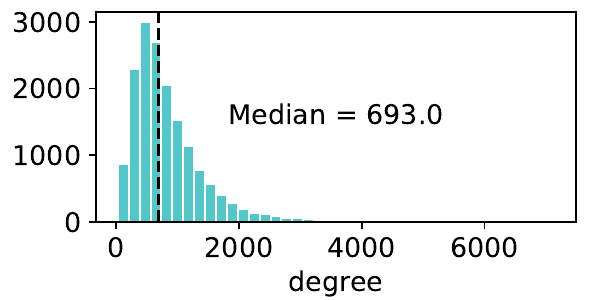}%\caption{fig1}
    }
    {\includegraphics[width=.24\textwidth]{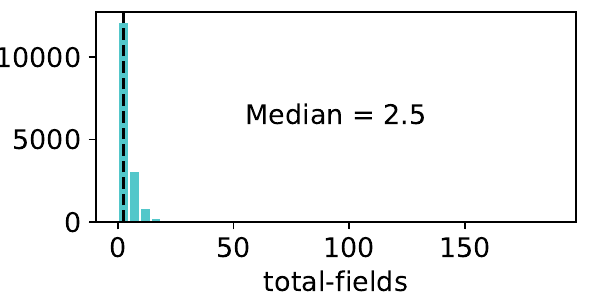}%\caption{fig1}
    }
    \quad
    \quad
    % \subfloat[The important features for InSep.]
    {\includegraphics[width=.24\textwidth]{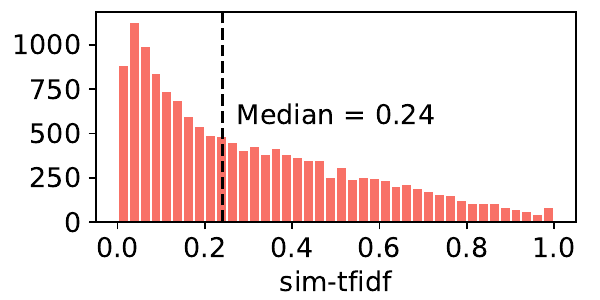}%\caption{fig1}
    }
    % \subfloat[The important features for InCol.]
    {\includegraphics[width=.24\textwidth]{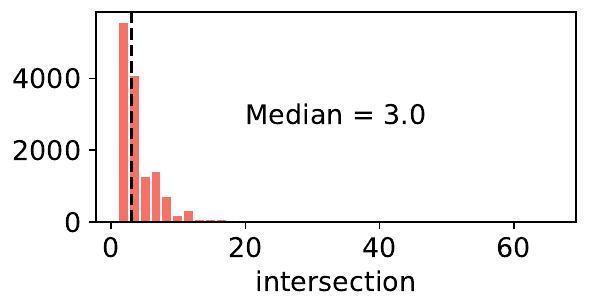}%\caption{fig1}
    }
    {\includegraphics[width=.24\textwidth]{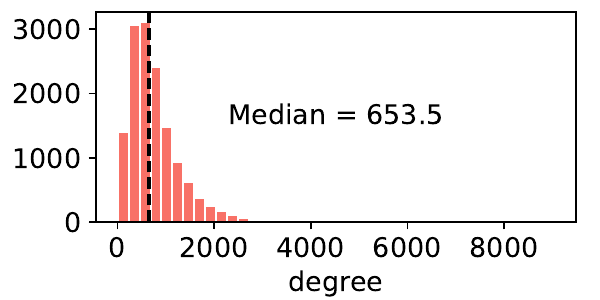}%\caption{fig1}
    }
    {\includegraphics[width=.24\textwidth]{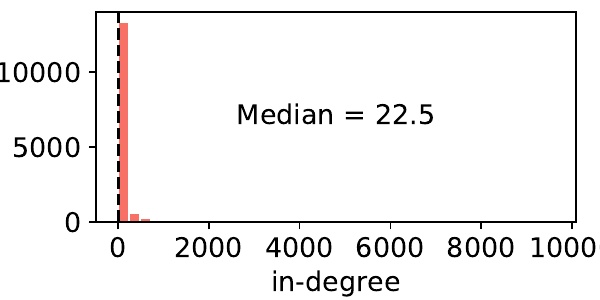}%\caption{fig1}
    }
    \caption{Distribution of features consistently ranked among the top 4 predictors for InSep (Blue) and InCol (Red)} 
    \label{fig: distributions}
\end{figure*}

We plot the distributions of the most important features to emphasize where the PDPs have more weight. 
Fig.~\ref{fig: distributions} shows the distributions of the features for predicting $\mathit{InSep}$ and $\mathit{InCol}$. For half of the separated pairs, each entity in the pair, on average, has fewer than 22.5 out-going dependencies, shares fewer than 693 common terms with other entities, and implements no more than two fields.
On the other hand, the medians of \emph{sim-tfidf}, \emph{intersection}, \emph{degree}, and \emph{in-degree} among collocated pairs are 0.24, 3, 653.5, and 22.5 respectively.
The distributions of values for important features are highly skewed, which should be considered when analyzing the impact of features. 
Table~\ref{tab: differences between smells and non-smells} further shows the median values of each important feature for smelly (\eg InSep) and non-smelly pairs (\eg Sep - InSep). 
We used the Mann-Whitney U test~\cite{mann1947test} to test whether the distributions of each feature for smelly and non-smelly pairs are equal. 
As can be seen, all features show significant differences; however, only \emph{out-degree} and \emph{degree} in separated pairs, and \emph{in-degree} in collocated pairs, exhibit small effect sizes. Similar distributions can be observed for the other two experiments.

\begin{figure*}
    \centering
    {\includegraphics[width=.24\textwidth]{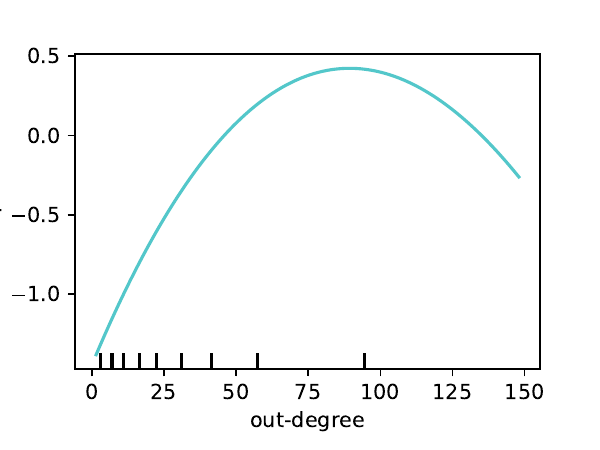}%\caption{fig1}
    }
    {\includegraphics[width=.24\textwidth]{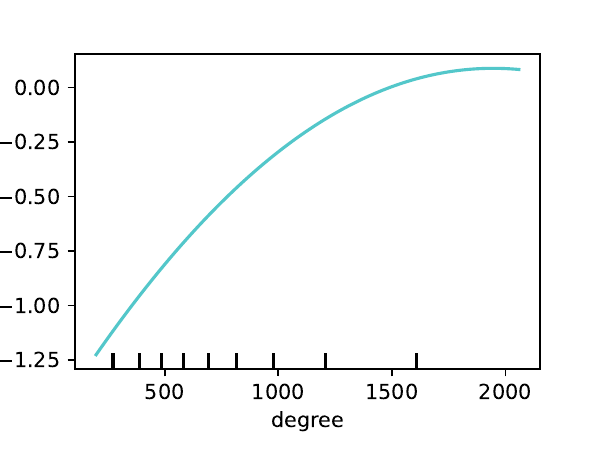}%\caption{fig1}
    }
    {\includegraphics[width=.24\textwidth]{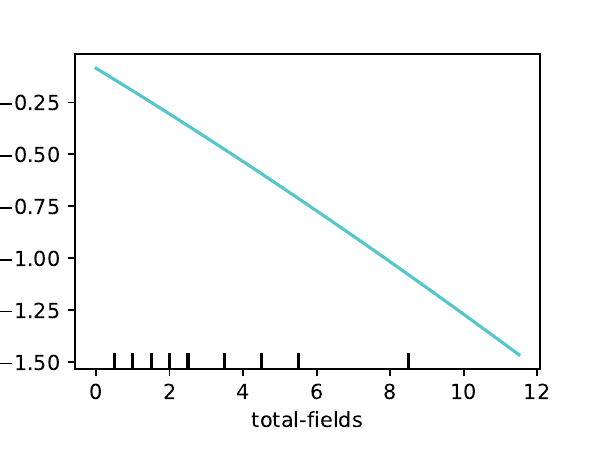}%\caption{fig1}
    }
    \quad
    \quad
    % \subfloat[The important features for InSep.]
    {\includegraphics[width=.24\textwidth]{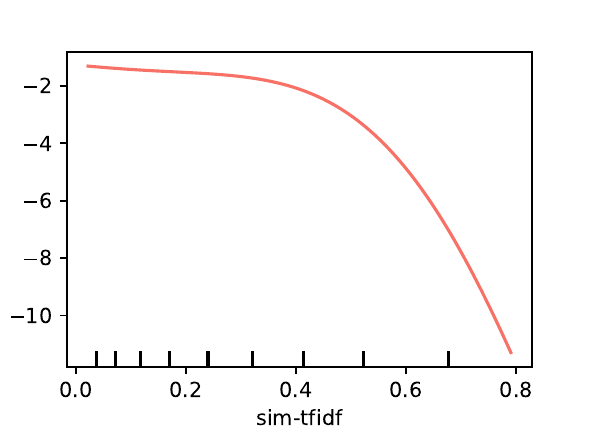}%\caption{fig1}
    }
    % \subfloat[The important features for InCol.]
    {\includegraphics[width=.24\textwidth]{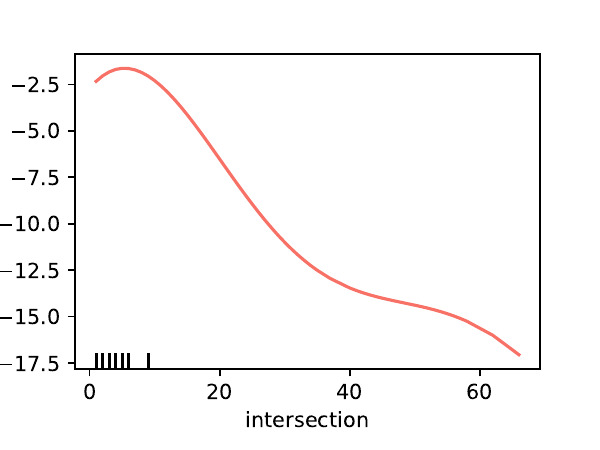}%\caption{fig1}
    }
    {\includegraphics[width=.24\textwidth]{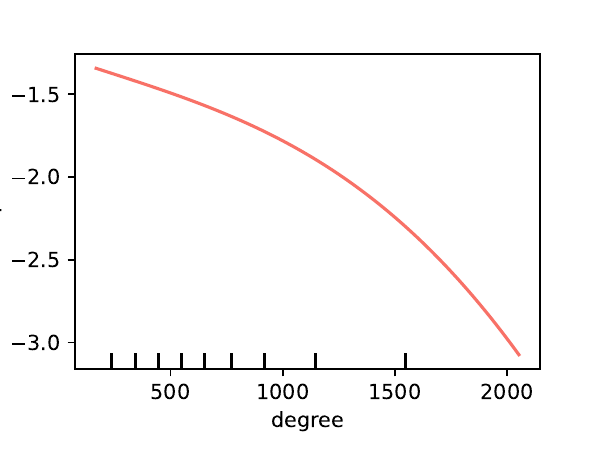}%\caption{fig1}
    }
    {\includegraphics[width=.24\textwidth]{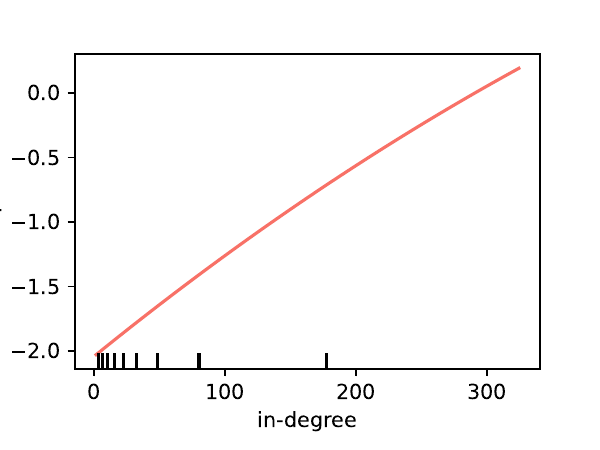}%\caption{fig1}
    }
    % }
    \caption{Partial dependence plots for each of the most important features for prediction InSep (Blue) and InCol (Red)} 
    \label{fig: partial dependence}
\end{figure*}

To visualize how a change in a pair characteristic (feature) impacts the models' decision-making for each class, we draw \emph{Partial Dependence Plots (PDP)}~\cite{molnar2020interpretable} for the most important features as presented in Fig.~\ref{fig: partial dependence}. PDP can visualize the marginal effect of a feature on the prediction of the machine learning model, highlighting linear, monotone or more complex relationships between the feature and the target.
In our cases, PDP can show how a change in a feature affects the models' likelihood to predict the occurrence of \emph{PairSmell}. The Y-axis represents the predicted probability for an instance belonging to \emph{PairSmell} ($\mathit{InSep}$ or $\mathit{InCol}$), and the tick marks on the X-axis represent the characteristic values.

\begin{table}
\caption{Medians for Smelly and Non-smelly Pairs Regarding the Most Important Pair Characteristics}
\label{tab: differences between smells and non-smells}
\scriptsize
\centering
\begin{tabular}{rrr r@{\hspace{0.001cm}} rrr}
\toprule
 \multicolumn{3}{c}{\textbf{Separated Pairs}} &  &\multicolumn{3}{c}{\textbf{Collocated Pairs}} \\
 \cmidrule{1-3} \cmidrule{5-7}
\textbf{Char.} & \textbf{InSep} & \textbf{Sep. - InSep} &  & \textbf{Char.} &  \textbf{InCol} & \textbf{Col. - InCol}  \\
\midrule
\emph{out-degree} *** & \cellcolor[HTML]{D9D9D9}41	& \cellcolor[HTML]{D9D9D9}19.5   & & \emph{sim-tfidf} *** & 0.09	& 0.26  \\
\emph{degree}*** & \cellcolor[HTML]{D9D9D9}941.5	& \cellcolor[HTML]{D9D9D9}647.5   && \emph{intersection}*** & 2 & 3  \\
\emph{total-fields}*** & 3	& 2.5   && \emph{degree}* & 703.5 & 649 \\
-   & - &   -   &&  in-degree***   & \cellcolor[HTML]{D9D9D9}99.5  & \cellcolor[HTML]{D9D9D9}19.5  \\
% -   & - &   -   &&  comment-density*  & 0.79  & 0.68  \\
\bottomrule
\end{tabular}
\begin{tablenotes}
        \item * ``Sep.'' and ``Col.'' denote separated pairs and collocated pairs, respectively.
        \item * $p < 0.1$, $p < 0.01$, and $p < 0.001$ are denoted by *, **, and ***, respectively.
        \item * \colorbox{mygray}{Gray} results show small effect size differences, while White results indicate an absolute difference smaller than 0.2.
\end{tablenotes}
\end{table}

For separated pairs, we see that as the number of outgoing edges (dependencies on other entities) of a pair increases, the probability of predicting the pair as $\mathit{InSep}$ first increases and then drops.
A pair with low \emph{out-degree} suggests that both entities do not rely much on other parts of the system -- they are self-contained or functionally independent of other entities and of each other. Thus, there is too little coupling to justify their collocation.  
In contrast, a pair with a high \emph{out-degree} (greater than 100, as shown in the figure) indicates that the two entities depend heavily on other specific entities within the system. Therefore, merging them into the same module could cause the module to over-rely on external modules, violating the principle of single responsibility.
% If we take into account the distribution in Fig.~\ref{fig: distributions}, since the median \emph{out-degree} is 22.5, the right portion of the plot in Fig.~\ref{fig: partial dependence} (\emph{out-degree} greater than 100) has less weight. That is, among the dataset, more pairs that are classified as $\mathit{Sep}$ due to low \emph{out-degree}.

\begin{tcolorbox}[boxsep=0pt,left=3pt,right=6pt,top=4pt,bottom=1pt] \itshape
\textbf{Finding \#4:} 
For separated pairs, the likelihood of predicting them as InSep initially rises and then falls with increasing outgoing edges. While low out-degree implies functional independence justifying separation, very high out-degree indicates excessive external dependencies that discourage merging to maintain separation of concerns among modules.
\end{tcolorbox}

For \emph{degree}, the partial dependence value increases monotonically with higher \emph{degree} values.
In other words, separated pairs with a higher \emph{degree} value are more likely to be classified as $\mathit{InSep}$.
This suggests that the more terms two entities share with other entities in the system, the more likely they are to be functionally relevant, thus they should be placed together rather than separated.

\begin{tcolorbox}[boxsep=0pt,left=3pt,right=6pt,top=4pt,bottom=1pt] \itshape
\textbf{Finding \#5:} 
For a separated pair, a higher \emph{degree} value suggests the two entities of the pair are more likely to be functionally relevant rather than separated (InSep). 
% \vspace*{0.5ex}
\end{tcolorbox}

\begin{table}
\caption{Percentages (\%) of Dependent and Non-dependent Separated Pairs Recognized as InSep, Divided by Different Number of Total Fields}
\label{tab: differences between dependent and non-dependent separated pairs}
\scriptsize
\centering
\begin{tabular}{rrrr}
\toprule
 & \textbf{$0 \leq $\emph{total-fields} $ \leq 2$}  & \textbf{$2 < $\emph{total-fields} $ \leq 8$} & \textbf{$8 < $\emph{total-fields} }  \\
\midrule
\textbf{Dep.} & 65 & 54  & 54\\
\textbf{Non-Dep.} & 17 & 19  & 20\\
% -   & - &   -   &&  comment-density*  & 0.79  & 0.68  \\
\bottomrule
\end{tabular}
\begin{tablenotes}
        \item * ``Dep.'' and ``Non-Dep.'' denote pairs with and without direct dependencies between them, respectively.
\end{tablenotes}
\end{table}

For \emph{total-fields}, the partial dependence decreases monotonically as the value increases -- separated pairs with fewer \emph{total-fields} are more likely to be classified as $\mathit{InSep}$.
In object-oriented design, classes with few fields are often relatively simple, which means that they are not central to a business functionality. 
However, the number of fields seems not to be a strong predictor of $\mathit{InSep}$; its permutation importance is below 0.04 -- far lower than that of other two features in Fig.~\ref{fig: importance features} (a).
Further inspection shows that low total fields only increases the likelihood of \emph{InSep} when combined with other evidence, \eg direct dependencies.
As shown in Table~\ref{tab: differences between dependent and non-dependent separated pairs}, only directly dependent pairs with low field counts exhibit high \emph{InSep} predictions. 
This suggests that when two dependent entities each have few fields, separating them may be unjustified due to insufficient encapsulated responsibilities.

\begin{tcolorbox}[boxsep=0pt,left=3pt,right=6pt,top=4pt,bottom=1pt] \itshape
\textbf{Finding \#6:} 
For a separated pair, low \emph{total-fields} slightly increases the probability of $\mathit{InSep}$, but only when combined with direct dependencies, suggesting that simple dependent entities are better kept together.

% \vspace*{0.5ex}
\end{tcolorbox}

For collocated pairs, we observe that higher values in \emph{sim-tfidf}, \emph{intersection}, and \emph{degree} generally decrease the likelihood that a collocated pair will be predicted as \emph{InCol}, while higher \emph{in-degree} increases it.
Specifically, as the \emph{sim-tfidf} value increases, the partial dependence decreases slightly initially and then drops significantly (after around 0.4).
That is, when the corresponding two entities have a low \emph{sim-tfidf} (\ie they are textually dissimilar), they are likely to be collocated inappropriately. 
This kind of entity pairs actually dominates the dataset according to the distribution in Fig.~\ref{fig: distributions}.
However, two entities with a high \emph{sim-tfidf} may have implemented relevant functionalities, thus keeping them into one module follows the single responsibility principle (at the module level), reducing the probability of inducing \emph{InCol}.

Similarly, as the \emph{intersection} increases, the predicted probability of the corresponding pair as \emph{InCol} decreases monotonically (except a small peak in the beginning).
That is, as the two entities in a pair share more common terms, they are more likely to implement related responsibilities, making it less likely that their collocation is considered inappropriate.

\begin{tcolorbox}[boxsep=0pt,left=3pt,right=6pt,top=4pt,bottom=1pt] \itshape
\textbf{Finding \#7:} 
For a collocated pair, a higher relatedness (indicated by \emph{sim-tfidf} or \emph{intersection}) suggests that the two entities potentially share more responsibilities or couplings, making it less likely to be affected by InCol. 
\end{tcolorbox}

The partial dependence line for \emph{degree} shows a clear downward trend as \emph{degree} increases.
A higher \emph{degree} suggests that the corresponding pair shares more common terms with others in the system, suggesting that the two entities likely belong to similar conceptual or functional areas, making their collocation more likely to be appropriate (\ie lower probability of being $\mathit{InCol}$).
In contrast, a lower \emph{degree} means the entities are not jointly connected to others, \ie they are more independent or specialized.

\begin{tcolorbox}[boxsep=0pt,left=3pt,right=6pt,top=4pt,bottom=1pt] \itshape
\textbf{Finding \#8:} 
For a collocated pair, a higher \emph{degree} value suggests that the two entities are less independent, decreasing the probability that it is affected by InCol. 
% \vspace*{0.5ex}
\end{tcolorbox}

For \emph{in-degree}, the partial dependence increases monotonically with a higher number of ingoing edges.
That is, the collocation of a pair with more ingoing edges is more likely to be identified as inappropriate (which leads to $\mathit{InCol}$).
According to Almugrin et al.~\cite{almugrin2016using}, more dependencies from others on an entity suggests more responsibilities it has in the system. 
Therefore, when a pair has more ingoing edges, it implies that the corresponding entities are responsible for more functionalities. Collocating such entities could potentially overload the module, which may raise concerns about the suitability of the collocation. Libraries and util packages are an exception to this, as we discuss in more detail in Section \ref{sec: limitationsOfPairSmell}.

\begin{tcolorbox}[boxsep=0pt,left=3pt,right=6pt,top=4pt,bottom=1pt] \itshape
\textbf{Finding \#9:} 
For a collocated pair, a higher \emph{in-degree} value indicates that the two entities have implemented many responsibilities, making the collocation less appropriate.
% \vspace*{0.5ex}
\end{tcolorbox}

% Finally, we see that as \emph{comment-density} increases, the probability of predicting the pair as $\mathit{InCol}$ first increases and then drops. 
% A pair with low \emph{comment-density} indicates that the code of both entities is not complex to understand, thus their collocation does not burden the current module. In contrast, a pair with a high \emph{comment-density} (greater than 2.5 from the figure) suggests that both entities implement complex functionalities, possibly the core functionality of the current module. 

% \begin{tcolorbox}[boxsep=0pt,left=3pt,right=6pt,top=4pt,bottom=1pt] \itshape
% \textbf{Finding \#2:} 
% For a collocated pair, the likelihood of predicting it as InCol first rises and then falls with increasing density of comments. While low comment-density implies simple code (encouraging collocation), dense comments indicate complex code (possibly the core functionality of the module).
% \end{tcolorbox}

\section{Illustrative Examples}
\label{sec: illustrative examples}

The interpretation of our models indicates that \emph{out-degree}, \emph{degree}, and \emph{total-fields} are strong indicators of $\mathit{InSep}$, while \emph{sim-tfidf}, \emph{intersection}, \emph{degree}, and \emph{in-degree} are strong indicators of $\mathit{InCol}$.
To gain deeper insights into these relationships, we manually analyze and illustrate representative examples from the dataset. Specifically, we present distinct example pairs for each type of \emph{PairSmell} -- $\mathit{InSep}$ and $\mathit{InCol}$ -- and the corresponding important features from the dataset.
Note that due to the similarity between the \emph{sim-tfidf} and \emph{intersection} features for $\mathit{InCol}$ pairs, we use a single example to illustrate both.

\noindent
\textbf{Example for InSep: out-degree.} 

\noindent
In the Hadoop project, two entities \texttt{LocatedBlocks.java} and \texttt{NamenodeFsck.java} are placed in separated modules.
However, this pair is detected as an instance of $\mathit{InSep}$, since all modularization tools consistently clustered them into the same module.

\begin{figure}
  \centering
  \includegraphics[width=0.8\linewidth]{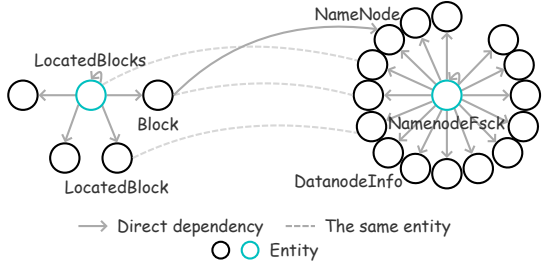}
  \caption{The outgoing dependencies of two separated entities, \texttt{LocatedBlocks.java} and \texttt{NamenodeFsck.java}, which share three connected entities. Moreover, there are tight couplings among the out-going dependencies.}
  \label{fig: insep_outdegree}
\end{figure}

We notice that the \emph{out-degree} value calculated for this pair is relatively high, at 89, with their outgoing dependencies illustrated in Fig.~\ref{fig: insep_outdegree}. 
Specifically, \texttt{LocatedBlocks.java} has outgoing dependencies on 5 entities, while \texttt{NamenodeFsck.java} depends on 16 entities.
Note that we omit the weight of these dependencies for clarity.
A high \emph{out-degree} for a separated pair often indicates that the entities they depend on are likely closely coupled.
For example, \texttt{Block.java}, which is depended by \texttt{LocatedBlocks.java}, further invokes \texttt{NameNode.java}, which is in turn depended upon by \texttt{NamenodeFsck.java}.
% Similarly, \texttt{DatanodeInfo.java}, which \texttt{NamenodeFsck} depends on, also relies on \texttt{LocatedBlock.java}, a dependency of \texttt{LocatedBlocks.java}.
This suggests tight couplings between the entities depended upon by the pair, which in turn lead to coupling between \texttt{LocatedBlocks.java} and \texttt{NamenodeFsck.java} themselves. 
Moreover, a high \emph{out-degree} increases the likelihood that the two entities share common dependencies, further reinforcing their coupling.
In this case, several dependencies of \texttt{LocatedBlocks.java}, such as \texttt{Block} and \texttt{LocatedBlock}, are also dependencies of \texttt{NamenodeFsck.java}.
Such tight couplings may question the appropriateness of placing them in separated modules.

\noindent
\textbf{Example for InSep: degree.} 

\noindent
In the Iotdb project, \texttt{TimeSeriesMetadataCache.java} and \texttt{MManager.java} constitute an instance of $\mathit{InSep}$, according to the identification results of \emph{PairSmell}. 

\begin{figure}
  \centering
  \includegraphics[width=0.76\linewidth]{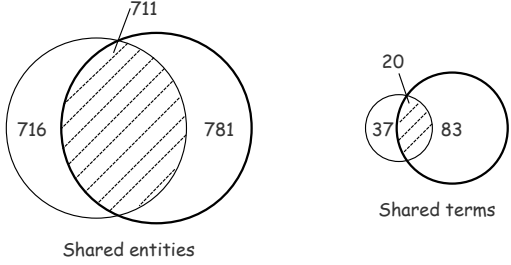}
  \caption{Overlap between \texttt{TimeSeriesMetadataCache.java} (left circle in both diagrams) and \texttt{MManager.java} (right circle in both diagrams) regarding shared entities (left diagram) and shared common terms (right diagram).
  }
  \label{fig: insep_degree}
\end{figure}

The \emph{degree} value of this separated pair is high at 3,100, suggesting that the two entities collectively share a total of 6200 common terms with other entities in the system (note that some terms may be shared with multiple entities). Specifically, \texttt{TimeSeriesMetadataCache.java} shares 2,311 common terms with 716 entities, while \texttt{MManager.java} shares 3,889 common terms with 781 entities.
A high \emph{degree} implies that both entities are conceptually connected to many others, indicating a high possibility of overlapping responsibilities between themselves. This raises questions about whether the separation between the two entities is appropriate. 
As shown in the left Venn diagram in Fig.~\ref{fig: insep_degree}, there are 711 entities that share common terms with both \texttt{TimeSeriesMetadataCache.java} and \texttt{MManager.java}, reinforcing their conceptual closeness.
In addition, the large number of common terms shared with other entities also suggests that the two entities may directly share many terms themselves. 
Indeed, as illustrated in the right Venn diagram, \texttt{TimeSeriesMetadataCache.java} and \texttt{MManager.java} share 20 common terms, including \texttt{logger}, \texttt{metadata}, and \texttt{measurement}, suggesting functional relevance between them. 
Functionally, \texttt{TimeSeriesMetadataCache.java} is responsible for caching \texttt{TimeSeriesMetadata} into database, while \texttt{MManager.java} handles serialization of all metadata info --including \texttt{TimeSeriesMetadata} --into files. 
This overlap in data handling responsibilities further supports their conceptual coupling.

\noindent
\textbf{Example for InSep: total-fields.} 

\noindent
In the Mahout project, \texttt{AggregatorMapper.java} and \texttt{TopKStringPatterns.java} are two code files from different modules, however, this separation is considered inappropriate.

\begin{figure}
  \centering
  \includegraphics[width=0.8\linewidth]{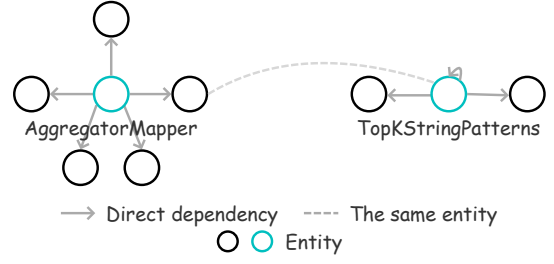}
  \caption{The outgoing~dependencies of two separated entities \\\texttt{AggregatorMapper.java}~and~\texttt{TopKStringPatterns.java} suggest they are structural dependent. 
  % Moreover, the low number of total fields suggest that they are relatively simple, which cannot justify their separation.
  % Both files declare zero fields and represent general concepts, thus their structural and semantic similarity may cast doubt on their separation.
  }
  \label{fig: insep_total_fields}
\end{figure}

A key observation is that these two entities, though located in separated modules, are closely related. 
As shown in Fig.~\ref{fig: insep_total_fields}, \texttt{AggregatorMapper.java} depends on \texttt{TopKStringPatterns.java} to define its input and output of one of its functions.
Both entities work together to identify and aggregate the top K frequent item patterns in a distributed dataset. 
Notably, the average total number of fields for this pair is low -- 0.5 -- indicating that they implement relatively simple functionality.
Specifically, \texttt{AggregatorMapper.java} groups patterns under individual items to facilitate top K computation, while \texttt{TopKStringPatterns.java} serves as a lightweight container for frequent patterns.
Given their structural dependencies and functional alignment, keeping them in separate modules might not be justified, as each encapsulates only simple and related responsibilities.

% the average total number of fields for this pair is low at 0.5, with only \texttt{TopKStringPatterns.java} implementing one field.

% A key observation is that both files declare zero fields. This absence of internal state indicates that they are designed to represent general concepts (rather than encapsulate detailed information), which likely implies a structural and semantic similarity between them.
% In fact, both files define custom exception classes that model specific error conditions -- \texttt{UnknownServiceException.java} for unknown services and \texttt{BadAuthException.java} for authentication failures.
% The design of minimal internal states aligns with their role as simple error indicators.
% Close inspection of the content of the two files (as shown in Fig.~\ref{fig: insep_total_fields}) reveals that both files were put in the same namespace, \texttt{org.apache.hadoop.hbase.ipc}, and inherit from the same base class, \texttt{FatalConnectionException}. Taken together, these factors—particularly the lack of fields—underscore their conceptual alignment and provide evidence that separating them into different components may not be justified.

\noindent
\textbf{Example for InCol: sim-tfidf and intersection.}

\noindent
In Pulsar, \texttt{Backoff.java} and \texttt{ConsumerBase.java} were clustered into one module, which produces an instance of \emph{InCol}.

\begin{figure}
  \centering
  \includegraphics[width=0.9\linewidth]{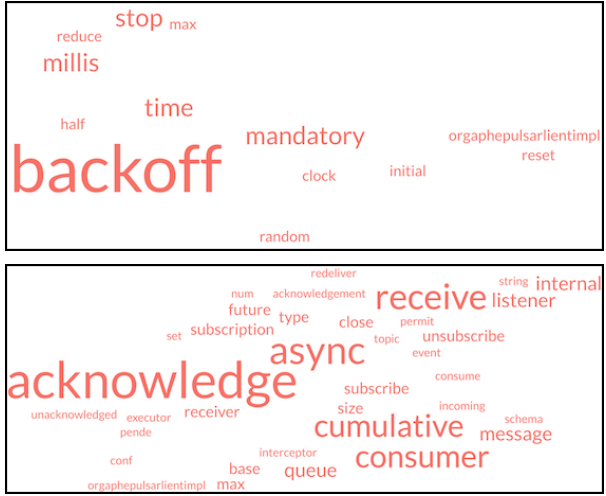}
  \caption{Word clouds generated for \texttt{Backoff.java} (up) and \texttt{ConsumerBase.java} (down), sharing only three terms.
  }
  \label{fig: incol_intersection}
\end{figure}

Our analysis finds that the \emph{sim-tfidf} value between the two entities is only 0.01 and they share only three terms in common (\emph{intersection} equals to 3)-- both indicating a low level of semantic relatedness.
Although each class shares common terms with many other entities (461 for \texttt{Backoff.java} and 815 for \texttt{ConsumerBase.java}), suggesting that they both handle broad responsibilities, the minimal overlap between them implies a weak direct connection between themselves.
The loose coupling between them is further illustrated in Fig.~\ref{fig: incol_intersection}.
From a functional standpoint, the two classes serve distinct purposes: \texttt{Backoff.java} handles exponential backoff logic, while \texttt{ConsumerBase.java} provides a foundational implementation for Pulsar consumers. These findings suggest that grouping them into the same module may not be justified.

\noindent
\textbf{Example for InCol: degree.}

\noindent
In Iotdb, two entities \texttt{BrokerAuthenticator.java} and \texttt{PayloadFormatter.java} are collocated in the \texttt{mqtt} module. However, this collocation might not be appropriate, which yields an instance of \emph{InCol}.

\begin{figure}
  \centering
  \includegraphics[width=0.64\linewidth]{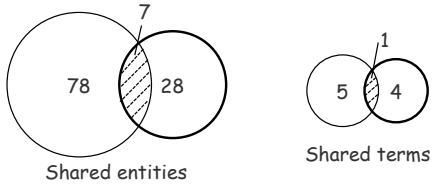}
  \caption{Overlap between \texttt{BrokerAuthenticator.java} (left circles in both diagrams) and \texttt{PayloadFormatter.java} (right circles in both diagrams) regarding shared entities and shared common terms.
  }
  \label{fig: incol_degree}
\end{figure}

The value of \emph{degree} for this collocated pair is low at 53.5, which corresponds to a high predicted probability of \emph{InCol} in Fig.~\ref{fig: partial dependence}.
Specifically, \texttt{BrokerAuthenticator.java} shares 79 common terms with 78 other entities, while \texttt{PayloadFormatter.java} shares 28 common terms with 20 entities. A low \emph{degree} suggests that both entities are only loosely connected to the rest of the system, implying they are likely functionally independent or highly specialized.
This raise questions about the appropriateness of their collocation. 
Although 7 entities share terms with both \texttt{BrokerAuthenticator.java} and \texttt{PayloadFormatter.java} (as shown in the left Venn diagram of Fig.~\ref{fig: incol_degree}), the two files themselves share only a single common term (as illustrated in the right Venn diagram). This reinforces their conceptual separation.
Indeed, their functionalities seem to be unrelated: \texttt{BrokerAuthenticator.java} validates client authentication, whereas \texttt{PayloadFormatter.java} formats a payload into a list of message.

% focuses on computing optimal Bloom filter parameters to reduce false positives, whereas \texttt{PayloadFormatter.java} provides utility methods for parsing XML files.

\noindent
\textbf{Example for InCol: in-degree.}

\noindent
\texttt{ApiKeys.java} and \texttt{Protocol.java} constitute a collocated pair within the \texttt{Protocol} module in Kafka. However, this pair is classified as $\mathit{InCol}$.

\begin{figure}
  \centering
  \includegraphics[width=0.8\linewidth]{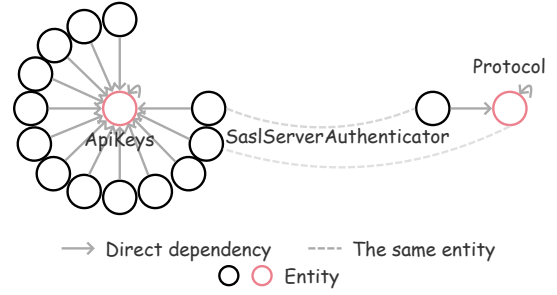}
  \caption{The ingoing dependencies of two collocated entities, \texttt{ApiKeys.java} and \texttt{Protocol.java}, within the \texttt{Protocol} module, indicating that \texttt{ApiKeys.java} has implemented many responsibilities.}
  \label{fig: incol_indegree}
\end{figure}

The \emph{in-degree} value of this collocated pair is high at 348, suggesting that the two entities have many in-going dependencies from others. The in-going dependencies are presented in Fig.~\ref{fig: incol_indegree}, which shows that \texttt{ApiKeys.java} has ingoing dependencies from 14 entities, while 2 entities (including itself) depend on \texttt{Protocol.java}.
A high \emph{in-degree} for a pair suggests that the corresponding entities might have implemented many responsibilities. 
In our case, although the \texttt{ApiKeys.java} entity was clustered into the module of \texttt{Protocol.java} (called \texttt{Protocol}), the entity itself has been invoked by 14 entities including 11 from other modules, suggesting that \texttt{ApiKeys.java} is coupled with many other modules. This indicates that clustering \texttt{ApiKeys.java} into the \texttt{Protocol} module might not be appropriate.
In addition, while \texttt{ApiKeys.java} enumerates high-level API keys, the \texttt{Protocol} module defines concrete API schema for all requests and responses.
Changes in API schema do not require changes to how APIs are enumerated, making it more appropriate to separate \texttt{ApiKeys.java} into its own module so that the two responsibilities are separate.

\section{Discussion}
\label{sec: discussion}
This section discusses how our findings would contribute to deepen the understanding of \emph{PairSmell}, especially its explainability, management, and potential limitations.

\subsection{Why PairSmell Is a Smell}

\textit{``A (code) smell is a surface indication that usually corresponds to a deeper problem in the system,''} according to the definition from Martin Fowler and Kent Beck~\cite{fowler2018refactoring, fowler2006code}.
This definition expresses two subtle points of the concept of smells.
First, a smell refers to a \emph{sniffable} problem that indicates the underlying inappropriate decision.
For example, \emph{Long Method} is recognizable by checking how many lines of code a method contains.
The second point is that, although smells can be promising indicators, they do not always pinpoint a real problem. 
For instance, some long methods are just fine. 
Smells that are problematic can negatively impact \emph{lifecycle properties} of a system, particularly maintainability~\cite{mumtaz2021systematic}, thus being regarded opportunities for refactoring in subsequent development.
In contrast, smells that do not generate maintenance ``interest'' are not true problems (or ``technical debt'')~\cite{xiao2021detecting}.
As Martin Fowler notes, the ``best'' smells are those that most of the time lead to really interesting problems.

In the case of \emph{PairSmell}, it is designed to be \emph{sniffable} -- as it arises when the MR of a pair violates the apt MR.
Thus, developers can recognize instances of \emph{PairSmell} once the apt MR of a separated pair is collocated ($\mathit{InSep}$) or the apt MR of a collocated pair is separated ($\mathit{InCol}$).
With the aid of the tooling introduced in our previous work~\cite{zhong2025pairsmell}, developers can efficiently and automatically detect all instances of \emph{PairSmell} within a system.

Regarding impact on maintainability, previous empirical evidence shows the negative influences of \emph{PairSmell} statistically.
To recap, entities in $\mathit{InSep}$ MRs co-change 190\% more than in other separated pairs, and entities in $\mathit{InCol}$ MRs co-change 35\% less than other collocated pairs, which dramatically deviates from well-structured modules. 
In this study, we advance the understanding by explaining \textit{why a pair exhibits \emph{PairSmell}} through the lens of pair characteristics. 
We disclose in Sections~\ref{sec: results} and \ref{sec: illustrative examples} that a pair affected by $\mathit{InSep}$ often exhibits at least one of three pair characteristics, and a pair affected by $\mathit{InCol}$ often manifests at least one of four pair characteristics.
Regarding the impact, Table~\ref{tab: characteristics and impacts} further discusses \textit{why pairs that manifest such characteristics are detrimental}, based on both our findings from this study and findings from studies on maintainability (\eg~\cite{sjoberg2012quantifying,yamashita2012code,jin2023dependency}).

\begin{table*}[tbh!]
\caption{Pair Characteristics, Corresponding Impacts, and Mitigation Means}
\label{tab: characteristics and impacts}
\scriptsize
\centering
\begin{tabular}{@{}lp{210pt}p{230pt}@{}}
\toprule
% \toprule
\textbf{Forms} & \textbf{Pair Characteristics: Potential Impacts} & \textbf{Possible Mitigation Means}\\ 
\midrule
\multirow{2}{*}{InSep} & \makecell[l]{\tabitem \parbox[t]{205pt}{\textit{High \textbf{out-degree}}: Two separated entities with many out-going edges may share dependencies, making changes to those shared dependents propagating to both entities in different modules.} \\
\tabitem  \parbox[t]{205pt}{\textit{High \textbf{degree}}: If two separated entities share many terms (or functionalities) with other parts of the system, they are likely overlapping responsibilities between modules and require cross-module changes when updating a feature.} \\
\tabitem  \parbox[t]{205pt}{\textit{Low \textbf{total-fields} (if dependent)}: When two dependent but separated entities each have a small number of fields, they tend to represent similar, simple responsibilities that are scattered across modules, making the overall system harder to understand.} \\
} & \makecell[l]{\tabitem \parbox[t]{205pt}{Depending on many low-level implementations can lead to high \textbf{out-degree}. To avoid this, developers can abstract interface and let the separated pairs depend on the interface (following the Dependency Inversion Principle).}  \\
\tabitem  \parbox[t]{205pt}{To reduce \textbf{degree}, developers can minimize the externally visible vocabularies of both entities by limiting public fields in accordance with the Information Hiding principle.}  \\
\tabitem  \parbox[t]{205pt}{To increase \textbf{total-fields}, developers can identify domain invariants within each entity and defined them as fields, strengthening object responsibility.}  \\
}\\
\midrule
\multirow{2}{*}{InCol} &  
\makecell[l]{\tabitem \parbox[t]{205pt}{\textit{Low \textbf{sim-tfidf}}: If two collocated entities are unrelated, it indicates that the current module handles distinct responsibilities, which can reduce its reusability.} \\
\tabitem  \parbox[t]{205pt}{\textit{Low \textbf{intersection}}: As with sim-tfidf, but here the relatedness is measured by intersection.} \\
\tabitem \parbox[t]{205pt}{\textit{Low \textbf{degree}}: When two collocated entities have limited overlap in terms (or functionalities) with other entities in the system, they are likely independent responsibilities within a single module, thereby reducing the maintainability of the current module.} \\
\tabitem \parbox[t]{205pt}{\textit{High \textbf{in-degree}}: When two collocated entities each have many in-going dependencies, it may indicate that they implement multiple responsibilities, thereby overloading the current module.} \\
} & \makecell[l]{\tabitem \parbox[t]{205pt}{To increase \textbf{sim-tfidf}, developers can replace entity-specific Data Transfer Objects with shared immutable value objects (to increase common identifiers).}   \\
\tabitem  \parbox[t]{205pt}{Local identifiers of the two entities can be renamed carefully to align terminology and reduce synonym fragmentation (policy and rule).}   \\
\tabitem \parbox[t]{205pt}{To increase \textbf{degree}, developers should use the same terms to represent the same concept within one module, following the pattern of Ubiquitous Language.}   \\
\tabitem \parbox[t]{205pt}{Both entities of the collocated pair serve multiple unrelated use cases and attract many dependents. Developers can apply Extract Class to further separate responsibilities apart.} \\
}\\
\bottomrule
\end{tabular}
% \begin{tablenotes}
%         \item * ``T.'' Stands for Themes.
% \end{tablenotes}
\end{table*}

\subsection{Implications for Modularity Research}

This section discusses implications for future research, particularly on research within the field of modularity.

\noindent
\textit{\textbf{Software modularization tools should carefully select pair characteristics and eliminate redundant characteristics.}}
Existing modularization studies often use multiple features between software entities (\aka pair characteristics) as the basis for analysis to derive well-designed modularization solutions.
For example, Eski and Buzluca~\cite{eski2018automatic} used various forms of structural coupling (\eg inheritance, aggregation, and method invocation) to extract microservices from a monolithic application.
However, our analysis reveals that the multiple features employed in the current community often exhibit collinear relationships (\eg \emph{dependency} and \emph{str-distance}), indicating that some features share significant similarities.
Integrating multiple similar features simultaneously into a modularization tool requires collecting redundant information, which leads to unnecessary resource consumption and offers a contrived analysis that may confuse developers regarding the real modularity issues in their system.
Furthermore, incorporating multiple similar features into a modularization algorithm (\eg clustering) may cause the resulting modularization to be dominated by a subset of primary features~\cite{rostami2021novel}. Therefore, we suggest that future research should explicitly account for feature similarity when designing modularization tools and carefully select features accordingly (\eg by combining Collinear analysis with ablation experiments).

\noindent
\textit{\textbf{In addition to pair \emph{relatedness}, modularization studies may also emphasize pair \emph{independence} and \emph{size} to improve the rationality of the resulting solutions.}}
Most existing modularization tools are designed around relationships between software entities (\eg~\cite{jin2019service,kalia2021mono2micro,zhong2025eism}), such as structural dependencies, semantic similarity, and co-change relationships.
However, some researchers argue that modularization solutions based solely on these explicit dependencies are often unsatisfactory, as they overlook additional insights that are important in architectural design (\eg implicit knowledge)~\cite{zhong2024refactoring}.
Our research found that, beyond relationships between entities, pair independence and size are also significantly associated with the modular design of a given pair (Section~\ref{sec: RQ3}).
Inspired by this finding, pair independence and size may help reveal relevant implicit knowledge, and future modularization research could place greater emphasis on these features during the design process, thereby producing more rational modularization solutions.
For example, pair independence reflects the relationship between a pair of entities and their broader system context. If two entities are highly dependent upon by many other entities within the system, aggregating them into a single module may be inappropriate, as it could result in modules with excessively large responsibilities (Section~\ref{sec: RQ3} and Section~\ref{sec: illustrative examples}).

\noindent
\textit{\textbf{The pair characteristics that are significantly associated with InSep (and InCol) pairs can be leveraged to assess modular relationships, complementing existing modularity metrics.}}
In terms of modularity metrics, most prior studies (\eg~\cite{jin2019service,kalia2021mono2micro}) regard high cohesion and loose coupling as the essential characteristics of `good' modules. For example, the widely used Modularization Quality (MQ)~\cite{mancoridis1998using} metric captures the extent to which entities within a module are tightly related while entities across different modules are loosely related. In contrast, only a small number of researchers have proposed alternative perspectives~\cite{candela2016using}.
Our findings may provide new insights for researchers into modularity measurement, thereby enriching existing approaches to assessing modularity. 
For two separated entities, the rationality of their separation can be evaluated using indicators such as outgoing dependencies, terms shared with other entities, and the number of declared fields within the pair (Section~\ref{sec: RQ3}). 
For example, by examining the number of outgoing dependencies of two entities, researchers can assess the extent to which the entities depend on other parts of the system, which may help justify their separation. 
Conversely, for two collocated entities, the fitness of their modular relationship can be assessed from multiple perspectives, including semantic similarity based on tf-idf, terms shared within the others, and in-going dependencies.

\subsection{Implications for Refactoring PairSmell}

\begin{figure}[b]
  \centering
  \includegraphics[width=0.99\linewidth]{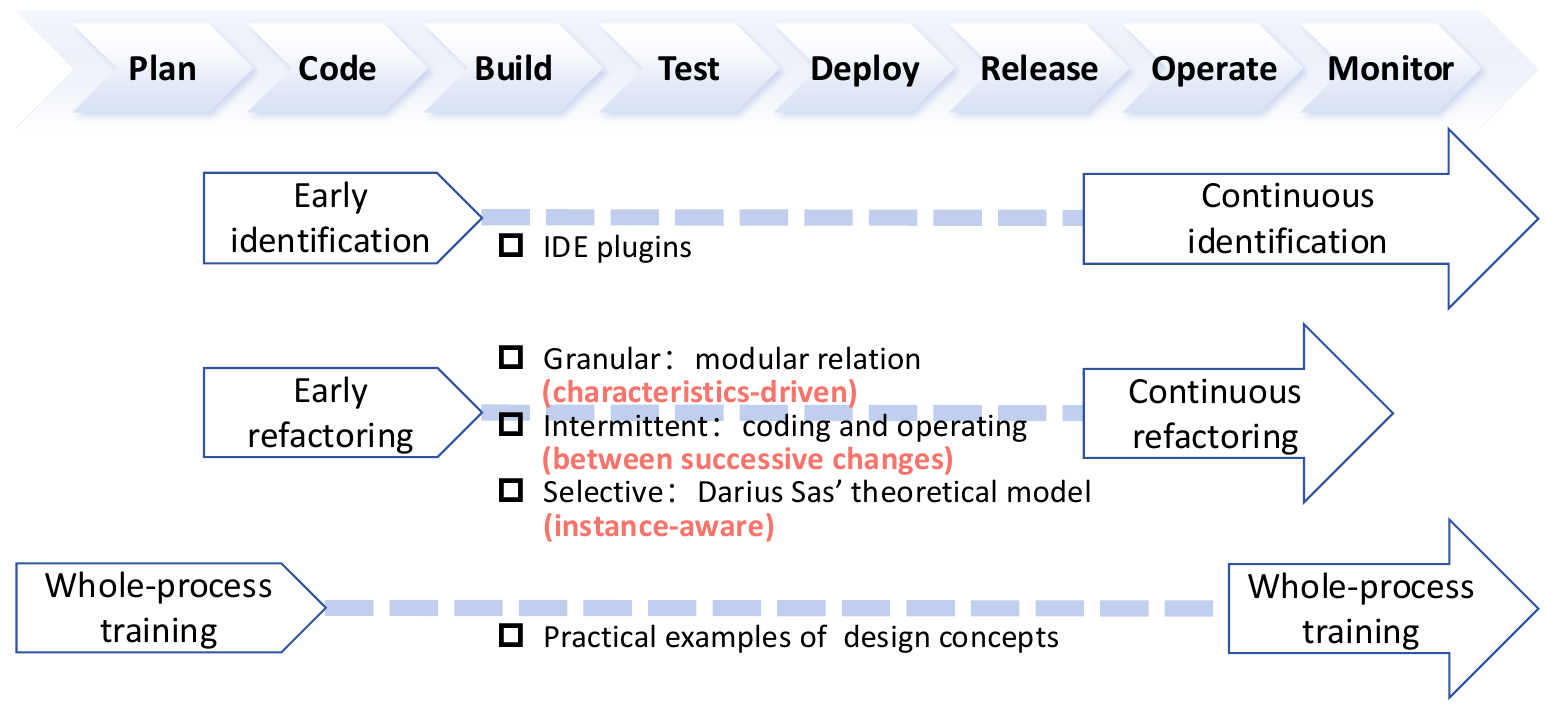}
  \caption{Management of PairSmell during the DevOps development process as adapted from prior work~\cite{zhong2025pairsmell}. Additions in ``\protect\redtext{}{Red}'' text are further insights obtained from this study.}
  \label{fig: lifecycle}
\end{figure}

In our previous study, we discussed how \emph{PairSmells} can be managed, through \textit{(1) early and continuous identification}, \textit{(2) granular, intermittent, and selective refactoring}, and \textit{(3) whole-process modular training}, as illustrated in Fig.~\ref{fig: lifecycle}. This subsection aims to enrich \emph{PairSmell} management strategies with the findings of this study (see \protect\redtext{}{red text} of Fig.~\ref{fig: lifecycle}).
% Note that the figure has been adapted with the insights obtained from this study.

To fix \emph{PairSmell}, typical refactoring strategies may be used: for an $\mathit{InSep}$ pair, developers could choose a single module to house both entities; while for an $\mathit{InCol}$ pair, developers could separate the current module to establish clear boundaries.
In addition, the close associations between pair characteristics and \emph{PairSmells} suggest that developers may be able to remove some instances of \emph{PairSmells} simply by adjusting the values of those important characteristics.
The detailed mechanisms for mitigating \emph{PairSmells} by leveraging these characteristics are exemplified in Table~\ref{tab: characteristics and impacts}.
% For example, developers may reduce the number of outgoing edges (\emph{out-degree}) from a separated pair to address an $\mathit{InSep}$ instance, \eg defining interfaces that abstract the dependencies and making the separated pair depend on the interfaces, following the Dependency Inversion Principle.
% For $\mathit{InCol}$, developers can increase the number of common terms shared by two entities (\emph{intersection}) in the same module, \eg use the same terms to represent the same concept, following the pattern of Ubiquitous Language~\cite{10495888}. 
This characteristic-driven strategy also helps developers make informed decisions during refactoring.
Developers can rely on the important pair characteristics to determine whether to perform a refactoring, by considering whether such a change may induce another instance of \emph{PairSmell}.
Regarding \emph{the time of refactor}, it is recommended to adopt \emph{intermittent floss refactoring} throughout the development lifecycle.
Based on this study, we further recommend performing such refactorings during the intervals between successive code changes (\eg between two sprints), as code changes may affect key pair characteristics and potentially influence the outcomes of \emph{PairSmell} analysis.
Regarding \emph{which smell to refactor first}, it is suggested to balance the severity of each smell instance and the remediation effort~\cite{sas2023architectural}.
Based on a more precise understanding of \emph{PairSmell}, we further suggest to incorporate specific refactoring strategies tailored to \emph{PairSmell} instances at hand (Table~\ref{tab: characteristics and impacts}).
For instance, developers might prioritize refactoring solutions that are easier to modify,
\eg revising terms used in a class (related to \emph{intersection}) is often more straightforward than reducing cross-module dependencies (related to \emph{out-degree}).

% that such a refactoring should be performed during the breaks of a series of code changes. 
% This is because the changes made may influence the values of those important pair characteristcis, which in turn may change the \emph{PairSmell} results.

% \textbf{3) Whole-process modular training.}
% Given the widespread occurrence of \emph{PairSmells} in numerous projects, our previous study suggests improving training for software design concepts using practical examples of modular smells such as \emph{PairSmells}.
% The illustrative examples in Section~\ref{sec: illustrative examples} play such a role in modular training. 
% For instance, ``Example for InSep: out-degree'' shows how two entities of separated modules are closely coupled via a series of direct and indirect structural dependencies, undermining the independence of the corresponding modules. Therefore, merging them into a single module might be an appropriate option. 

\subsection{Limitations of PairSmell}

From this study, we observed two limitations of \emph{PairSmell}.
\label{sec: limitationsOfPairSmell}

\begin{itemize}
    \item \textit{The usage of lowest-level folders as the actual modular structure may lead to false positives.} As discussed in our previous study, the folder structure of a software repository can be represented by a tree hierarchy of folders and sub-folders. In this study, we notice that developers may create sub-folders inside a java package for organizational purposes, which leads to several sub-modules inside the jave module.
    Consequently, \emph{PairSmell} instances detected using the lowest-level folders might not be problematic at the package level. For instance, two entities separated into two lowest-level folders may still belong to the same package and thus be considered collocated at the package level -- leading to false positives of $\mathit{InSep}$. 
    This suggests that developers can tailor the definition of the actual `modules' when detecting \emph{PairSmells} to suit their specific context (\eg using java \emph{packages} or the \emph{module} system introduced in Java 9).

    \item \textit{Utility packages may need to be excluded when analyzing %modular structures such as when inspecting 
    PairSmell.}
    Utility packages typically contain reusable helper classes that provide general-purpose functionalities (\eg exception handling), rather than cohesive responsibilities.
    This can lead to false positives of $\mathit{InCol}$.
    For instance, in Cassandra, \texttt{BloomCalculations.java} and \texttt{XMLUtils.java} are two entities collocated in the \texttt{utils} module despite being unrelated.
    Thus, excluding utility packages can improve the accuracy of \emph{PairSmell} detection.

\end{itemize}

\section{Threats to Validity}
\label{sec: ttv}

This section presents potential threats to the validity of our study and means we used to mitigate them.

\noindent
\textbf{Construct Validity.}
The construct validity might be influenced by two factors. On the one hand, the 19 features used to train our model might not capture significant pair characteristics, raising concerns about their appropriateness and comprehensiveness.
For example, the co-change between two entities reflects their evolutionary relationship, which has not been included in our model. This feature may explain the appropriateness of a modular relation for a pair.
However, such information cannot be easily collected for each pair in the development environment, and our aim is to provide insights to aid developers in inspecting modular design during development.
To mitigate this threat, we selected a set of 19 features that span six dimensions of pair characteristics by consulting the literature through a rapid review.
In addition, we carefully adjust certain characteristics to ensure their suitability (\ie consistency between different themes).
We believe these features are suitable, as we cross-referenced them with both modular design decisions identified in the literature and empirical characteristics observed in code repositories.
On the other hand, the AURPC values of our models are low in some cases, which may appear to limit their practical usefulness in imbalanced settings.
However, AUPRC is strongly affected by extreme class imbalance, as reflected by the best baseline AUPRC of 0.262. Compared with these baselines, our model improves AUPRC by up to 177\%, indicating a substantially improved ability to prioritize positive instances.
More importantly, our model is not intended to function as a standalone classifier, but to expose and explain indicators of \emph{PairSmell}.
Consequently, predictive metrics such as AUPRC do not fully capture the model’s explanatory utility, a positioning that is consistent with prior work using machine learning models primarily for explanation~\cite{weeraddana2024characterizing}.

\noindent
\textbf{Internal Validity.}
The methodology designed in our study can influence internal validity with two possible threats.
First, the validity of our dataset may be affected by several factors.
To mitigate potential risks, we built our dataset based on a previously published dataset.
To avoid noise, we excluded pairs in which at least one of the entities is a test file, configuration file, or inner class. 
To address the class imbalance issue, we applied Inverse Transform Sampling~\cite{devroye1986sample} to generate samples from the entire population, which may affect the distribution of our dataset. 
To ensure that the sampled dataset remains representative, we determined the size of our sample using a 99\% confidence level and a 1\% margin of error.
Second, our investigation in this study may not be comprehensive enough.
To address the potential threat, we conducted a rapid literature review to extract and synthesize significant pair characteristics from 178 primary studies.
Our final set of 19 features encompasses six key aspects: relatedness, distance, independence, complexity, cohesion, and size.
In addition, the pair characteristics obtained were used as independent variables in our analysis to investigate those that are most indicative of \emph{PairSmell}. 
Additionally, we complemented our quantitative analysis with qualitative analysis by exploring and presenting multiple examples to illustrate and support our findings.

\noindent
\textbf{External Validity.}
The generalization and application of our findings can be impacted by two threats.
First, the findings of our rapid review may not be generalizable enough to represent all significant pair characteristics that describe the relationships between two entities. To minimize this threat, we include primary studies from the most up-to-date and systematic reviews related to software modularization. Our review of the literature incorporates 178 primary studies, including 143 papers examining the clustering of software modules and 35 papers studying microservices-oriented decomposition.
Our review process rigorously followed the guideline of thematic analysis~\cite{Cruzes2011}.
Second, we are aware that our analysis of the relationships between pair characteristics and \emph{PairSmell} may not be generalizable to other datasets since all pairs were collected from open-source Java projects. 
To this end, multiple criteria were used to select projects with varying scales (283 to 1,456 entities), business domains (\eg database, search engine), and other characteristics (Table~\ref{tab: systems}).
Moreover, we argue that the pair characteristics we study are language-agnostic (\eg entity dependencies) as they are not strongly tied to Java language features. 
Finally, from the large-scale dataset of separated and collocated pairs, we performed the sampling process three times and conducted experiments on each of the samples, all of which yielded similar findings. Therefore, this threat has been mitigated, at least to some extent.

% To this end, a set of criteria were used to select projects with varying scales, business domains, and other characteristics.
% In addition, we performed the sampling process three times and conducted experiments on each of the samples, all of which yielded similar findings.

\section{Conclusions}
\label{sec:conclusions}

In order to characterize and explain \emph{PairSmell}, this study first conducts a rapid review on 178 primary studies related to software modularization to collect pair characteristics that can be used to describe the relationships between two entities. We then examine 6,135,877 distinct pairs from 11 open-source Java projects. 
Overall 19 pair characteristics from the review, spanning the relatedness, distance, independence, complexity, cohesion, and size aspects, are then analyzed using machine learning models to understand their impact on the occurrence of \emph{PairSmells}. 
Our findings highlight that the most influential factors for $\mathit{InSep}$ are out-going dependencies, terms
shared with others, and declared fields, and those for  $\mathit{InCol}$ are semantic similarity based on tf-idf, terms shared between the pair, terms shared with others, and in-going dependencies.
We complement the quantitative analysis with a set of illustrative examples to investigate modular relations of pairs and understand how they are impacted by factors such as high \emph{out-degree}.
The findings of this study provides valuable insights not only for managing \emph{PairSmell}, but also for leveraging salient pair characteristics to support future research and practice in modular design.

\ifCLASSOPTIONcompsoc
  \section*{Acknowledgments}
\else
  \section*{Acknowledgment}
\fi

% This work is supported by the Key Research and Development Program of Jiangsu Province (No.BE2021002-2), the National Natural Science Foundation of China (No.62302210, No.62072227, No.62202219), the Natural Science Foundation of Jiangsu Province (No.BK20241195), and the Innovation Projects and Overseas Open Projects of State Key Laboratory for Novel Software Technology (Nanjing University) (ZZKT2024A18, ZZKT2024B07, KFKT2023A09, KFKT2023A10, KFKT2024A02, KFKT2024A13, KFKT2024A14).

This work is supported by the National Natural Science Foundation of China (No.62502211, No.62302210, No.62572237), 2024 Development and Testing Tools Project (CEIEC-2024-ZM02-0066), the Natural Science Foundation of Jiangsu Province (No.BK20241195), and the Innovation Project and Overseas Open Project of State Key Laboratory for Novel Software Technology (Nanjing University) (KFKT2025A19, KFKT2025A17, KFKT2025A20, ZZKT2026A10, ZZKT2026A37, ZZKT2026A40, ZZKT2026A44, ZZKT2026A50, ZZKT2026A52, KFKT2026A11, KFKT2026A16, KFKT2026A18, KFKT2024A02, KFKT2024A13, KFKT2024A14, KFKT2026B43).

\bibliographystyle{IEEEtran}
\bibliography{references}

@inproceedings{wang2024reposvul,
  title={ReposVul: A Repository-Level High-Quality Vulnerability Dataset},
  author={Wang, Xinchen and Hu, Ruida and Gao, Cuiyun and Wen, Xin-Cheng and Chen, Yujia and Liao, Qing},
  booktitle={Proceedings of the 2024 IEEE/ACM 46th International Conference on Software Engineering: Companion Proceedings},
  pages={472--483},
  year={2024}
}

@article{mitchell2006automatic,
  title={On the automatic modularization of software systems using the bunch tool},
  author={Mitchell, Brian S and Mancoridis, Spiros},
  journal={IEEE Transactions on Software Engineering},
  volume={32},
  number={3},
  pages={193--208},
  year={2006},
  publisher={IEEE}
}

@inproceedings{garcia2009identifying,
  title={Identifying architectural bad smells},
  author={Garcia, Joshua and Popescu, Daniel and Edwards, George and Medvidovic, Nenad},
  booktitle={Proceedings of the 13th European Conference on Software Maintenance and Reengineering},
  pages={255--258},
  year={2009},
  organization={IEEE}
}

@article{sjoberg2012quantifying,
  title={Quantifying the effect of code smells on maintenance effort},
  author={Sj{\o}berg, Dag IK and Yamashita, Aiko and Anda, Bente CD and Mockus, Audris and Dyb{\aa}, Tore},
  journal={IEEE Transactions on Software Engineering},
  volume={39},
  number={8},
  pages={1144--1156},
  year={2012},
  publisher={IEEE}
}

@inproceedings{jin2023dependency,
  title={Dependency facade: The coupling and conflicts between android framework and its customization},
  author={Jin, Wuxia and Dai, Yitong and Zheng, Jianguo and Qu, Yu and Fan, Ming and Huang, Zhenyu and Huang, Dezhi and Liu, Ting},
  booktitle={Proceedings of the IEEE/ACM 45th International Conference on Software Engineering},
  pages={1674--1686},
  year={2023},
  organization={IEEE}
}

@inproceedings{yamashita2012code,
  title={Do code smells reflect important maintainability aspects?},
  author={Yamashita, Aiko and Moonen, Leon},
  booktitle={Proceedings of the 28th IEEE International Conference on Software Maintenance},
  pages={306--315},
  year={2012},
  organization={IEEE}
}

@inproceedings{mancoridis1998using,
  title={Using automatic clustering to produce high-level system organizations of source code},
  author={Mancoridis, Spiros and Mitchell, Brian S and Rorres, Chris and Chen, Y and Gansner, Emden R},
  booktitle={Proceedings of 6th International Workshop on Program Comprehension},
  pages={45--52},
  year={1998},
  organization={IEEE}
}

@inproceedings{wong2011detecting,
  title={Detecting software modularity violations},
  author={Wong, Sunny and Cai, Yuanfang and Kim, Miryung and Dalton, Michael},
  booktitle={Proceedings of the 33rd International Conference on Software Engineering},
  pages={411--420},
  year={2011}
}

@article{mann1947test,
  title={On a test of whether one of two random variables is stochastically larger than the other},
  author={Mann, Henry B and Whitney, Donald R},
  journal={The Annals of Mathematical Statistics},
  pages={50--60},
  year={1947},
  publisher={JSTOR}
}

@article{sas2023architectural,
  title={An architectural technical debt index based on machine learning and architectural smells},
  author={Sas, Darius and Avgeriou, Paris},
  journal={IEEE Transactions on Software Engineering},
  volume={49},
  number={8},
  pages={4169--4195},
  year={2023},
  publisher={IEEE}
}

@inproceedings{Mo2015,
  title        = {Hotspot patterns: The formal definition and automatic detection of architecture smells},
  author       = {Mo, Ran and Cai, Yuanfang and Kazman, Rick and Xiao, Lu},
  booktitle    = {Proceedings of the 12th Working IEEE/IFIP Conference on Software Architecture},
  pages        = {51--60},
  year         = {2015},
  organization = {IEEE}
}

@inproceedings{le2018empirical,
  title={An empirical study of architectural decay in open-source software},
  author={Le, Duc Minh and Link, Daniel and Shahbazian, Arman and Medvidovic, Nenad},
  booktitle={Proceedings of the 15th International Conference on Software Architecture},
  pages={176--17609},
  year={2018},
  organization={IEEE}
}

@book{shull2008guide,
  title={Guide to advanced empirical software engineering},
  author={Shull, Forrest and Singer, Janice and Sj{\o}berg, Dag IK},
  volume={93},
  year={2008},
  publisher={Springer}
}

@article{al2012precise,
  title={A precise method-method interaction-based cohesion metric for object-oriented classes},
  author={Al Dallal, Jehad and Briand, Lionel C},
  journal={ACM Transactions on Software Engineering and Methodology},
  volume={21},
  number={2},
  pages={1--34},
  year={2012},
  publisher={Association for Computing Machinery}
}

@article{saito2015precision,
  title={The precision-recall plot is more informative than the ROC plot when evaluating binary classifiers on imbalanced datasets},
  author={Saito, Takaya and Rehmsmeier, Marc},
  journal={PloS one},
  volume={10},
  number={3},
  pages={e0118432},
  year={2015},
  publisher={Public Library of Science San Francisco, CA USA}
}

@article{almugrin2016using,
  title={Using indirect coupling metrics to predict package maintainability and testability},
  author={Almugrin, Saleh and Albattah, Waleed and Melton, Austin},
  journal={Journal of Systems and Software},
  volume={121},
  pages={298--310},
  year={2016},
  publisher={Elsevier}
}

@inproceedings{arvanitou2015introducing,
  title={Introducing a ripple effect measure: A theoretical and empirical validation},
  author={Arvanitou, Elvira-Maria and Ampatzoglou, Apostolos and Chatzigeorgiou, Alexander and Avgeriou, Paris},
  booktitle={Proceedings of the 9th International Symposium on Empirical Software Engineering and Measurement},
  pages={1--10},
  year={2015},
  organization={IEEE}
}

@article{athanasopoulos2014cohesion,
  title={Cohesion-driven decomposition of service interfaces without access to source code},
  author={Athanasopoulos, Dionysis and Zarras, Apostolos V and Miskos, George and Issarny, Valerie and Vassiliadis, Panos},
  journal={IEEE Transactions on Services Computing},
  volume={8},
  number={4},
  pages={550--562},
  year={2014},
  publisher={IEEE}
}

@article{glorie2009splitting,
  title={Splitting a large software repository for easing future software evolution—an industrial experience report},
  author={Glorie, Marco and Zaidman, Andy and van Deursen, Arie and Hofland, Lennart},
  journal={Journal of Software Maintenance and Evolution: Research and Practice},
  volume={21},
  number={2},
  pages={113--141},
  year={2009},
  publisher={Wiley Online Library}
}

@article{tricco2015scoping,
  title={A scoping review of rapid review methods},
  author={Tricco, Andrea C and Antony, Jesmin and Zarin, Wasifa and Strifler, Lisa and Ghassemi, Marco and Ivory, John and Perrier, Laure and Hutton, Brian and Moher, David and Straus, Sharon E},
  journal={BMC Medicine},
  volume={13},
  pages={1--15},
  year={2015},
  publisher={Springer}
}

@article{abgaz2023decomposition,
  title={Decomposition of Monolith Applications Into Microservices Architectures: A Systematic Review},
  author={Abgaz, Yalemisew and McCarren, Andrew and Elger, Peter and Solan, David and Lapuz, Neil and Bivol, Marin and Jackson, Glenn and Yilmaz, Murat and Buckley, Jim and Clarke, Paul},
  journal={IEEE Transactions on Software Engineering},
  year={2023},
  publisher={IEEE}
}

@article{zhong2025eism,
  title={EISM: An Interactive and Collaborative Approach for Software Modularization},
  author={Zhong, Chenxing and Li, Chao and Zhang, He},
  journal={Journal of Systems and Software},
  pages={112726},
  year={2025},
  publisher={Elsevier}
}

@book{baldwin2000design,
  title={Design rules: The power of modularity},
  author={Baldwin, Carliss Young and Clark, Kim B},
  volume={1},
  year={2000},
  publisher={MIT press}
}

@inproceedings{griffith2014design,
  title={Design pattern decay: The case for class grime},
  author={Griffith, Isaac and Izurieta, Clemente},
  booktitle={Proceedings of the 8th ACM/IEEE International Symposium on Empirical Software Engineering and Measurement},
  pages={1--4},
  year={2014},
  publisher={ACM}
}

@inproceedings{kruger2020empirical,
  title={An empirical analysis of the costs of clone-and platform-oriented software reuse},
  author={Kr{\"u}ger, Jacob and Berger, Thorsten},
  booktitle={Proceedings of the 28th ACM Joint Meeting on European Software Engineering Conference and Symposium on the Foundations of Software Engineering},
  pages={432--444},
  year={2020}
}

@inproceedings{Cruzes2011,
  title={Recommended steps for thematic synthesis in software engineering},
  author={Cruzes, Daniela S. and Dyba, Tore},
  booktitle={Proceedings of the 5th International Symposium on Empirical Software Engineering and Measurement},
  pages={275--284},
  year={2011},
  organization={IEEE}
}

@inproceedings{oyetoyan2015decision,
  title={A decision support system to refactor class cycles},
  author={Oyetoyan, Tosin Daniel and Cruzes, Daniela Soares and Thurmann-Nielsen, Christian},
  booktitle={Proceedings of the 31th IEEE International Conference on Software Maintenance and Evolution},
  pages={231--240},
  year={2015},
  organization={IEEE}
}

@book{molnar2020interpretable,
  title={Interpretable machine learning},
  author={Molnar, Christoph},
  year={2020},
  publisher={Lulu. com}
}

@article{jin2019service,
  title         = {Service candidate identification from monolithic systems based on execution traces},
  author        = {Jin, Wuxia and Liu, Ting and Cai, Yuanfang and Kazman, Rick and Mo, Ran and Zheng, Qinghua},
  journal       = {IEEE Transactions on Software Engineering},
  volume        ={47},
  number        ={5},
  pages         ={987--1007},
  year          = {2019},
  publisher     = {IEEE}
}

@article{wang2017automatic,
  title={Automatic software refactoring via weighted clustering in method-level networks},
  author={Wang, Ying and Yu, Hai and Zhu, Zhiliang and Zhang, Wei and Zhao, Yuli},
  journal={IEEE Transactions on Software Engineering},
  volume={44},
  number={3},
  pages={202--236},
  year={2017},
  publisher={IEEE}
}

@misc{depends2022,
  year = {2022},
  title =        {DEPENDS},
  howpublished = {\url{https://github.com/multilang-depends/depends}}
}

@inproceedings{devroye1986sample,
  title={Sample-based non-uniform random variate generation},
  author={Devroye, Luc},
  booktitle={Proceedings of the 18th Conference on Winter Simulation},
  pages={260--265},
  year={1986}
}

@misc{nltk2023,
  year = {2023},
  title =        {NLTK},
  howpublished = {\url{https://www.nltk.org/}}
}

@misc{scikit2024,
  year = {2024},
  title =        {Scikit-learn},
  howpublished = {\url{https://scikit-learn.org/stable/whats_new/v1.4.html#version-1-4-1}}
}

@misc{ctags2022,
  year = {2022},
  title =        {Ctags},
  howpublished = {\url{https://github.com/universal-ctags/ctags}}
}

@misc{cloc2025,
  year = {2024},
  title =        {CLOC},
  howpublished = {\url{https://github.com/AlDanial/cloc/commits/v2.00}}
}

@misc{ck2022,
  year = {2022},
  title =        {CK},
  howpublished = {\url{https://github.com/mauricioaniche/ck/releases/tag/ck-0.7.0}}
}

@inproceedings{kalia2021mono2micro,
  title={Mono2Micro: A practical and effective tool for decomposing monolithic Java applications to microservices},
  author={Kalia, Anup K and Xiao, Jin and Krishna, Rahul and Sinha, Saurabh and Vukovic, Maja and Banerjee, Debasish},
  booktitle={Proceedings of the 29th ACM Joint Meeting on European Software Engineering Conference and Symposium on the Foundations of Software Engineering},
  pages={1214--1224},
  year={2021}
}

@article{fred2005combining,
  title={Combining multiple clusterings using evidence accumulation},
  author={Fred, Ana LN and Jain, Anil K},
  journal={IEEE transactions on Pattern Analysis and Machine Intelligence},
  volume={27},
  number={6},
  pages={835--850},
  year={2005},
  publisher={IEEE}
}

@article{zhang2022weighted,
  title={Weighted clustering ensemble: A review},
  author={Zhang, Mimi},
  journal={Pattern Recognition},
  volume={124},
  pages={108428},
  year={2022},
  publisher={Elsevier}
}

@article{CAI2023107322,
title = {Software design analysis and technical debt management based on design rule theory},
journal = {Information and Software Technology},
volume = {164},
pages = {107322},
year = {2023},
issn = {0950-5849},
doi = {https://doi.org/10.1016/j.infsof.2023.107322},
url = {https://www.sciencedirect.com/science/article/pii/S0950584923001775},
author = {Cai, Yuanfang and Kazman, Rick}
}

@inproceedings{fontana2017arcan,
  title={Arcan: A tool for architectural smells detection},
  author={Fontana, Francesca Arcelli and Pigazzini, Ilaria and Roveda, Riccardo and Tamburri, Damian and Zanoni, Marco and Di Nitto, Elisabetta},
  booktitle={Proceedings of the 2017 IEEE International Conference on Software Architecture Workshops},
  pages={282--285},
  year={2017},
  organization={IEEE}
}

@inproceedings{mo2016decoupling,
  title={Decoupling level: A new metric for architectural maintenance complexity},
  author={Mo, Ran and Cai, Yuanfang and Kazman, Rick and Xiao, Lu and Feng, Qiong},
  booktitle={Proceedings of the 38th International Conference on Software Engineering},
  pages={499--510},
  year={2016},
  organization={IEEE}
}

@inproceedings{zhong2025pairsmell,
  title={PairSmell: A Novel Perspective Inspecting Software Modular Structure},
  author={Zhong, Chenxing and Feitosa, Daniel and Avgeriou, Paris and Huang, Huang and Li, Yue and Zhang, He},
  booktitle={Proceedings of the 47th International Conference on Software Engineering},
  pages={1--12},
  year={2025},
  organization={IEEE}
}

@inproceedings{tan2020first,
  title={A first look at good first issues on GitHub},
  author={Tan, Xin and Zhou, Minghui and Sun, Zeyu},
  booktitle={Proceedings of the 28th ACM Joint Meeting on European Software Engineering Conference and Symposium on the Foundations of Software Engineering},
  pages={398--409},
  year={2020}
}

@article{weeraddana2024characterizing,
  title={Characterizing timeout builds in continuous integration},
  author={Weeraddana, Nimmi and Alfadel, Mahmoud and McIntosh, Shane},
  journal={IEEE Transactions on Software Engineering},
  year={2024},
  publisher={IEEE}
}

@inproceedings{mcintosh2018fix,
  title={Are fix-inducing changes a moving target? A longitudinal case study of just-in-time defect prediction},
  author={McIntosh, Shane and Kamei, Yasutaka},
  booktitle={Proceedings of the 40th International Conference on Software Engineering},
  pages={560--560},
  year={2018}
}

@article{rostami2021novel,
  title={A novel community detection based genetic algorithm for feature selection},
  author={Rostami, Mehrdad and Berahmand, Kamal and Forouzandeh, Saman},
  journal={Journal of Big Data},
  volume={8},
  number={1},
  pages={2},
  year={2021},
  publisher={Springer}
}

@inproceedings{eski2018automatic,
  title={An automatic extraction approach: Transition to microservices architecture from monolithic application},
  author={Eski, Sinan and Buzluca, Feza},
  booktitle={Proceedings of the 19th International Conference on Agile Software Development: Companion},
  pages={1--6},
  year={2018}
}

@article{mumtaz2021systematic,
  title={A systematic mapping study on architectural smells detection},
  author={Mumtaz, Haris and Singh, Paramvir and Blincoe, Kelly},
  journal={Journal of Systems and Software},
  volume={173},
  pages={110885},
  year={2021},
  publisher={Elsevier}
}

@article{pourasghar2021graph,
  title={A graph-based clustering algorithm for software systems modularization},
  author={Pourasghar, Babak and Izadkhah, Habib and Isazadeh, Ayaz and Lotfi, Shahriar},
  journal={Information and Software Technology},
  volume={133},
  pages={106469},
  year={2021},
  publisher={Elsevier}
}

@book{newman2021building,
  title={Building microservices},
  author={Newman, Sam},
  year={2021},
  publisher={O'Reilly Media}
}

@article{mo2019architecture,
  title={Architecture anti-patterns: Automatically detectable violations of design principles},
  author={Mo, Ran and Cai, Yuanfang and Kazman, Rick and Xiao, Lu and Feng, Qiong},
  journal={IEEE Transactions on Software Engineering},
  volume={47},
  number={5},
  pages={1008--1028},
  year={2019},
  publisher={IEEE}
}

@book{fowler2018refactoring,
  title={Refactoring},
  author={Fowler, Martin},
  year={2018},
  publisher={Addison-Wesley Professional}
}

@book{spearman1961proof,
  title={The proof and measurement of association between two things.},
  author={Spearman, Charles},
  year={1961},
  publisher={Appleton-Century-Crofts}
}

@inproceedings{schroder2021search,
  title={Search-based software re-modularization: A case study at Adyen},
  author={Schr{\"o}der, Casper and van der Feltz, Adriaan and Panichella, Annibale and Aniche, Maur{\'\i}cio},
  booktitle={Proceedings of the 43rd International Conference on Software Engineering: Software Engineering in Practice},
  pages={81--90},
  year={2021},
  organization={IEEE}
}

@article{sarhan2020software,
  title={Software module clustering: An in-depth literature analysis},
  author={Sarhan, Qusay I and Ahmed, Bestoun S and Bures, Miroslav and Zamli, Kamal Z},
  journal={IEEE Transactions on Software Engineering},
  volume={48},
  number={6},
  pages={1905--1928},
  year={2022},
  publisher={IEEE}
}

@article{xiao2021detecting,
  title={Detecting the Locations and Predicting the Costs of Compound Architectural Debts},
  author={Xiao, Lu and Cai, Yuanfang and Kazman, Rick and Mo, Ran and Feng, Qiong},
  journal={IEEE Transactions on Software Engineering},
  volume    = {48},
  number    = {9},
  pages     = {3686--3715},
  year      = {2022},
  publisher={IEEE}
}

@article{biau2016random,
  title={A random forest guided tour},
  author={Biau, G{\'e}rard and Scornet, Erwan},
  journal={Test},
  volume={25},
  number={2},
  pages={197--227},
  year={2016},
  publisher={Springer}
}

@incollection{awad2008support,
  title={Support vector machines},
  author={Awad, Mamoun and Khan, Latifur},
  booktitle={Intelligent Information Technologies: Concepts, Methodologies, Tools, and Applications},
  pages={1138--1146},
  year={2008},
  publisher={IGI Global}
}

@book{hosmer2013applied,
  title={Applied logistic regression},
  author={Hosmer Jr, David W and Lemeshow, Stanley and Sturdivant, Rodney X},
  year={2013},
  publisher={John Wiley \& Sons}
}

@inproceedings{boyd2013area,
  title={Area under the precision-recall curve: Point estimates and confidence intervals},
  author={Boyd, Kendrick and Eng, Kevin H and Page, C David},
  booktitle={Proceedings of the 2013 European Conference on Machine Learning and Knowledge Discovery in Databases},
  pages={451--466},
  year={2013},
  organization={Springer}
}

@article{hanley1982meaning,
  title={The meaning and use of the area under a receiver operating characteristic (ROC) curve},
  author={Hanley, James A and McNeil, Barbara J},
  journal={Radiology},
  volume={143},
  number={1},
  pages={29--36},
  year={1982}
}

@article{liu2024prevalence,
  title={Prevalence and severity of design anti-patterns in open source programs—A large-scale study},
  author={Liu, Alan and Lefever, Jason and Han, Yi and Cai, Yuanfang},
  journal={Information and Software Technology},
  volume={170},
  pages={107429},
  year={2024},
  publisher={Elsevier}
}

@article{quattrocchi2024cromlech,
  title={Cromlech: Semi-Automated Monolith Decomposition Into Microservices},
  author={Quattrocchi, Giovanni and Cocco, Davide and Staffa, Simone and Margara, Alessandro and Cugola, Gianpaolo},
  journal={IEEE Transactions on Services Computing},
  year={2024},
  publisher={IEEE}
}

@article{yang2022enhancing,
  title={Enhancing software modularization via semantic outliers filtration and label propagation},
  author={Yang, Kaiyuan and Wang, Junfeng and Fang, Zhiyang and Wu, Peng and Song, Zihua},
  journal={Information and Software Technology},
  volume={145},
  pages={106818},
  year={2022},
  publisher={Elsevier}
}

@article{zhong2024refactoring,
  title={Refactoring microservices to microservices in support of evolutionary design},
  author={Zhong, Chenxing and Li, Shanshan and Zhang, He and Huang, Huang and Yang, Lanxin and Cai, Yuanfang},
  journal={IEEE transactions on software engineering},
  year={2024},
  publisher={IEEE}
}

@article{candela2016using,
  title={Using cohesion and coupling for software remodularization: Is it enough?},
  author={Candela, Ivan and Bavota, Gabriele and Russo, Barbara and Oliveto, Rocco},
  journal={ACM Transactions on Software Engineering and Methodology},
  volume={25},
  number={3},
  pages={1--28},
  year={2016},
  publisher={ACM New York}
}

@article{zhong2023measuring,
  title={On measuring coupling between microservices},
  author={Zhong, Chenxing and Zhang, He and Li, Chao and Huang, Huang and Feitosa, Daniel},
  journal={Journal of Systems and Software},
  pages={111670},
  year={2023},
  publisher={Elsevier}
}

@article{fowler2006code,
  title={Code Smell},
  author={Fowler, Martin},
  URL={https://martinfowler.com/bliki/CodeSmell.html},
  year={2006}
}

@book{martin2018clean,
  title={Clean code: A craftsman’s guide to software structure and design},
  author={Martin, Robert C},
  year={2018},
  publisher={Pearson Education}
}

@article{teymourian2022fast,
  title={A fast clustering algorithm for modularization of large-scale software systems},
  author={Teymourian, Navid and Izadkhah, Habib and Isazadeh, Ayaz},
  journal={IEEE Transactions on Software Engineering},
  year={2022},
  volume={48},
  number={4},
  pages={1451--1462},
  publisher={IEEE}
}

@book{kitchenham2007guidelines,
  title={Guidelines for performing systematic literature reviews in software engineering},
  author={Kitchenham, Barbara and Charters, Stuart},
  year={2007},
  publisher={Keele, UK}
}

\vfill

\end{document}

\title{Appendix for Two Additional Experiments}
\author{Chenxing~Zhong, Daniel Feitosa, Paris Avgeriou, Huang~Huang, Wei~Song, and He~Zhang \\
\thanks{Manuscript created June, 2025; 
% This work was developed by the IEEE Publication Technology Department. This work is distributed under the \LaTeX \ Project Public License (LPPL) ( http://www.latex-project.org/ ) version 1.3. A copy of the LPPL, version 1.3, is included in the base \LaTeX \ documentation of all distributions of \LaTeX \ released 2003/12/01 or later. The opinions expressed here are entirely that of the author. No warranty is expressed or implied. User assumes all risk.
}}

\markboth{Journal of \LaTeX\ Class Files,~Vol.~18, No.~9, September~2020}%
{How to Use the IEEEtran \LaTeX \ Templates}

\maketitle

\begin{abstract}
This document describes the results for two additional experiments, including the plots of permutation importance and partial dependence. The format used to present the results is consistent with the main manuscript.

% most common article elements and how to use the IEEEtran class with \LaTeX \ to produce files that are suitable for submission to the Institute of Electrical and Electronics Engineers (IEEE).  IEEEtran can produce conference, journal and technical note (correspondence) papers with a suitable choice of class options.
\end{abstract}

% \begin{IEEEkeywords}
% Class, IEEEtran, \LaTeX, paper, style, template, typesetting.
% \end{IEEEkeywords}

\section{Data Collection and Sampling}

Our final dataset for the 2nd experiment consists of 16,541 separated pairs and 14,372 collocated pairs spanning 11 projects. Among these separated pairs, 3,111 are $\mathit{InSep}$ instances, and among the collocated pairs, 1,382 are $\mathit{InCol}$ instances.

Our final dataset for the 3rd experiment consists of 16,541 separated pairs and 14,372 collocated pairs spanning 11 projects. Among these separated pairs, 3,105 are $\mathit{InSep}$ instances, and among the collocated pairs, 1,378 are $\mathit{InCol}$ instances.

\section{Can the pair characteristics be used as indicators for \emph{PairSmell}? (RQ2)}

Table~\ref{tab: performance for candidate models_2} and Table~\ref{tab: performance for candidate models_3} show the evaluation results regarding ROC-AUC and AUPRC for the 2nd and 3rd experiments, respectively. 
\change{R1.3}{Similar to the results for the 1st experiment (in the main manuscript), the ROC-AUC for the baseline models round up to 0.5, which is the expected ROC-AUC values when the model makes random predictions or always predicts the same class. 
Compared to baseline models, we can observe that SVM models in the 2nd experiment improve the ROC-AUC by 30.2\% (39.4\% in the 3rd experiment) for separated pairs and by 58.8\% (61.2\%) for collocated pairs. 
The consistently high ROC-AUC improvement in three experiments shows that the pair characteristics we study can be leveraged to predict if a pair is affected by PairSmell. In other words, they can be used as indicators of PairSmell.

Regarding the AUPRC metric, we see that SVM models yield in the 2nd experiment an AUPRC of 0.417 (0.465 in the 3rd experiment) for separated pairs and an AUPRC of 0.471 (0.478) for collocated pairs, surpassing the best baselines with an improvement of 1.59 (1.60) times and 2.91 (2.95) times, respectively. 
This results support the model's effectiveness in distinguishing positive instances and minimizing false positives, which is especially crucial in our dataset of separated and collocated pairs with an imbalanced class distribution.
}

\begin{table}
\caption{Comparison of performance for candidate models (the 2nd Experiment)}
\label{tab: performance for candidate models_2}
\scriptsize
\centering
\begin{tabular}{p{36pt}rr r@{\hspace{0.001cm}} rr}
\toprule
 & \multicolumn{2}{c}{\textbf{Separated Pairs}} &  &\multicolumn{2}{c}{\textbf{Collocated Pairs}} \\
 \cmidrule{2-3} \cmidrule{5-6}
\multirow{-2}{*}{\textbf{Models}} & \textbf{ROC-AUC} & \textbf{AUPRC} &  & \textbf{ROC-AUC} & \textbf{AUPRC}  \\
\midrule
LR & 0.652	& 0.418   && 0.772 & 0.386  \\
SVM & 0.651	& 0.417  & &0.794	& 0.471	  \\
RF & 0.636	& 0.413 && 0.774	& 0.609	  \\
\midrule
SB & 0.489	& 0.263  && 0.503	& 0.162  \\
CB & 0.5	& 0  && 0.5	& 0	  \\
TB & 0.5    & 0.188 &&  0.5 & 0.096 \\
\bottomrule
\end{tabular}
\begin{tablenotes}
        \item * ``LR'', ``SVM'', and ``RF'' represent logistic regression, support vector machine, and random forest, respectively.
        \item * ``SB'', ``CB'', and ``TB'' represent stratified baseline, constant baseline, and theoretical baseline, respectively.
\end{tablenotes}
\end{table}

\begin{table}
\caption{Comparison of performance for candidate models (the 2nd Experiment)}
\label{tab: performance for candidate models_3}
\scriptsize
\centering
\begin{tabular}{p{36pt}rr r@{\hspace{0.001cm}} rr}
\toprule
 & \multicolumn{2}{c}{\textbf{Separated Pairs}} &  &\multicolumn{2}{c}{\textbf{Collocated Pairs}} \\
 \cmidrule{2-3} \cmidrule{5-6}
\multirow{-2}{*}{\textbf{Models}} & \textbf{ROC-AUC} & \textbf{AUPRC} &  & \textbf{ROC-AUC} & \textbf{AUPRC}  \\
\midrule
LR & 0.664	& 0.431   && 0.794 & 0.404  \\
SVM & 0.697	& 0.465  & &0.806	& 0.478	  \\
RF & 0.635	& 0.412 && 0.789	& 0.61	  \\
\midrule
SB & 0.517	& 0.29  && 0.503	& 0.162  \\
CB & 0.5	& 0  && 0.5	& 0	  \\
TB & 0.5    & 0.188 &&  0.5 & 0.096 \\
\bottomrule
\end{tabular}
\begin{tablenotes}
        \item * ``LR'', ``SVM'', and ``RF'' represent logistic regression, support vector machine, and random forest, respectively.
        \item * ``SB'', ``CB'', and ``TB'' represent stratified baseline, constant baseline, and theoretical baseline, respectively.
\end{tablenotes}
\end{table}

\section{Which pair characteristics are the most important indicators for \emph{PairSmell}? (RQ3)}

\begin{figure*}[t]
    \centering
    \subfloat[The important features for predicting InSep]{\includegraphics[width=.5\textwidth]{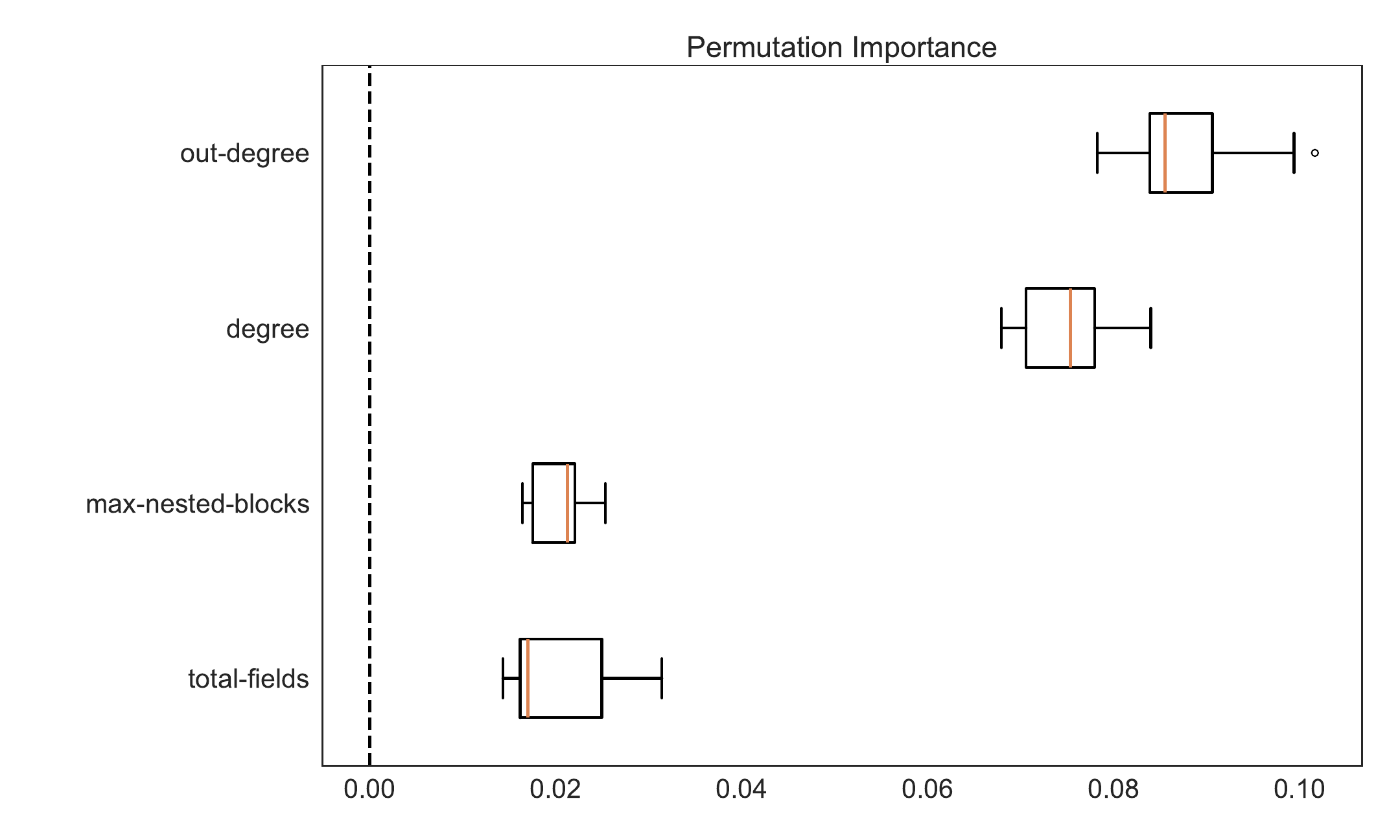}%\caption{fig1}
    }
    % \quad
    \subfloat[The important features for predicting InCol]{\includegraphics[width=.5\textwidth]{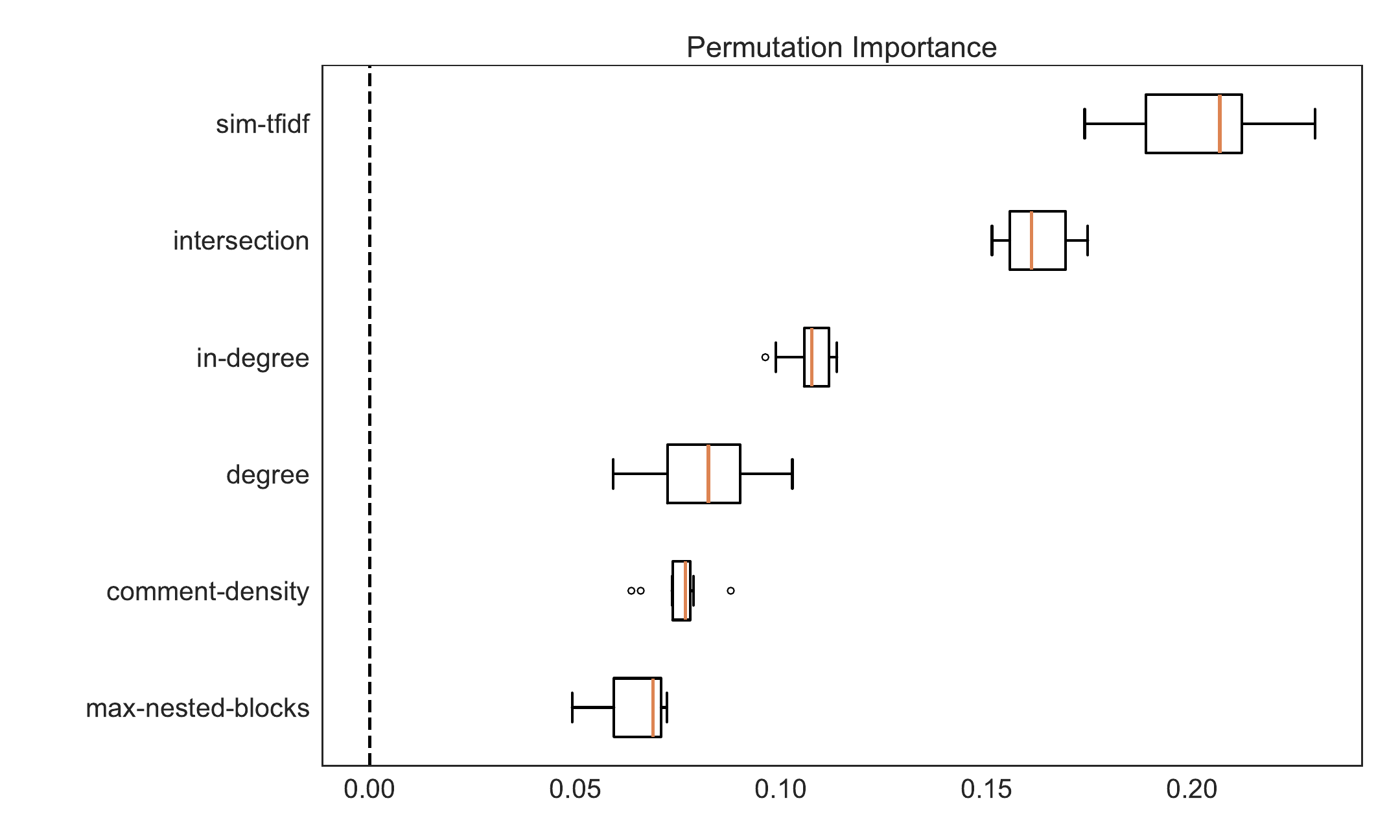}%\caption{fig1}
    }
    \quad
    \centering
    \caption{The most important features associated with the occurrence of InSep and InCol (The 2nd experiment)} 
    \label{fig: importance features_experiment2}
\end{figure*}

\begin{figure*}[t]
    \centering
    \subfloat[The important features for predicting InSep]{\includegraphics[width=.5\textwidth]{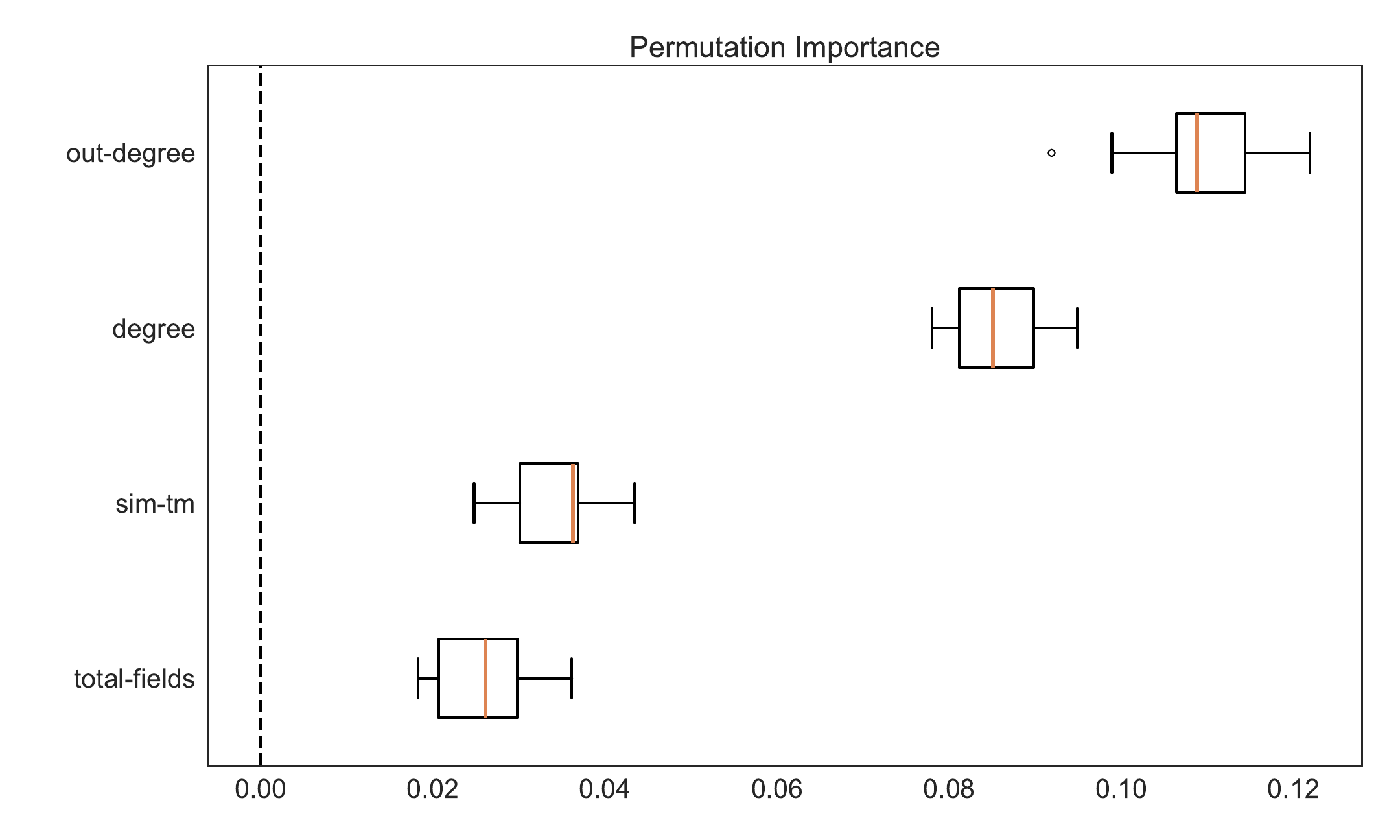}%\caption{fig1}
    }
    % \quad
    \subfloat[The important features for predicting InCol]{\includegraphics[width=.5\textwidth]{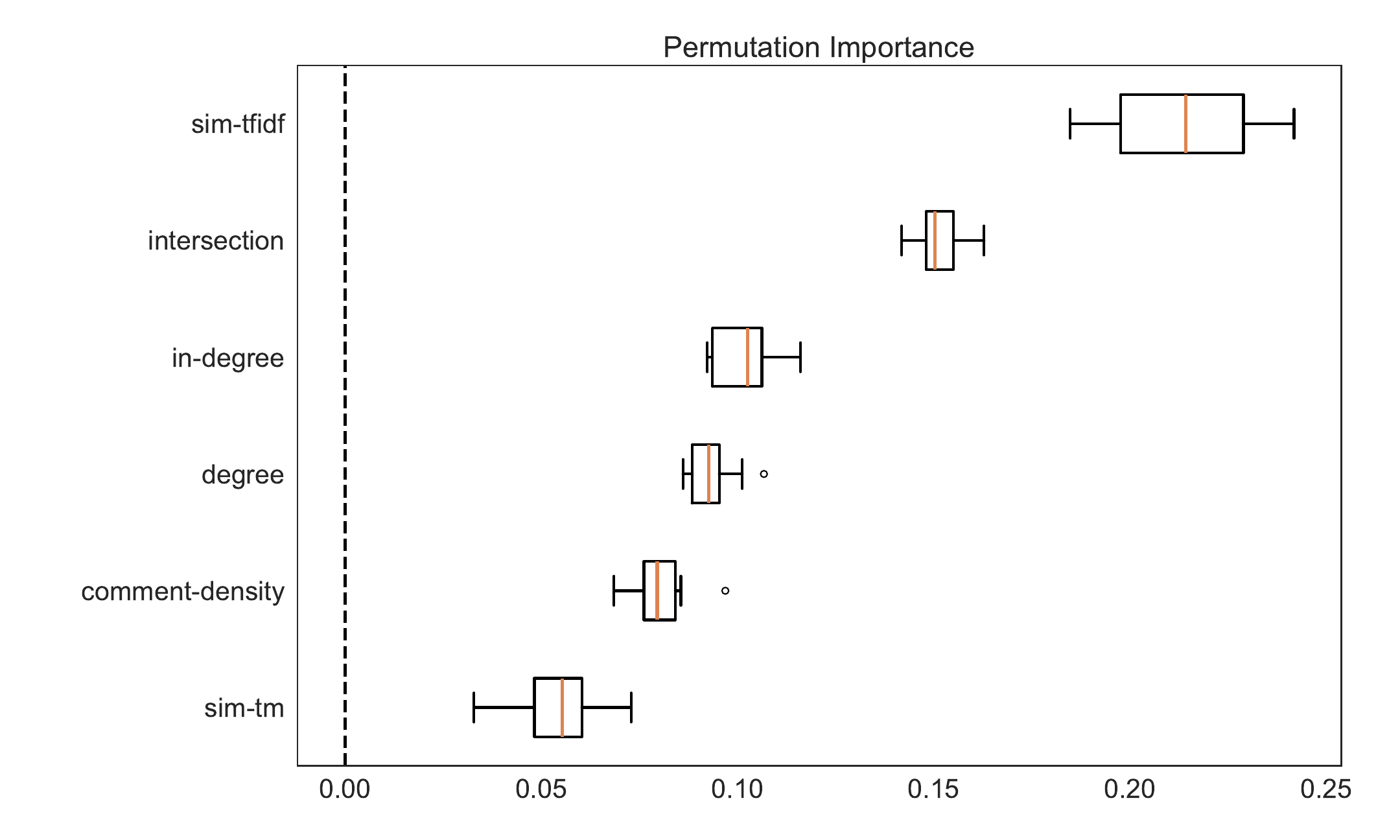}%\caption{fig1}
    }
    \quad
    \centering
    \caption{The most important features associated with the occurrence of InSep and InCol (The 3rd experiment)} 
    \label{fig: importance features_experiment3}
\end{figure*}

\change{R1.3}{Fig.~\ref{fig: importance features_experiment2} and Fig.~\ref{fig: importance features_experiment3} present the features whose permutation importance is higher than zero to predict the occurrence of $\mathit{InSep}$ and $\mathit{InCol}$, for the 2nd and 3rd experiments.
From both figures, we observed that \emph{out-degree}, \emph{degree}, and \emph{total-fields} are the most important indicators of the occurrence of $\mathit{InSep}$, as they consistently rank among the top 4 indicators.
Similarly, in terms of collocated pairs, \emph{sim-tfidf}, \emph{intersection}, \emph{degree}, and \emph{in-degree}, are the most important indicators of the occurrence of $\mathit{InCol}$ (as they rank among the top 4 indicators in all three experiments).}

\begin{figure*}[t]
    \centering
    {\includegraphics[width=.24\textwidth]{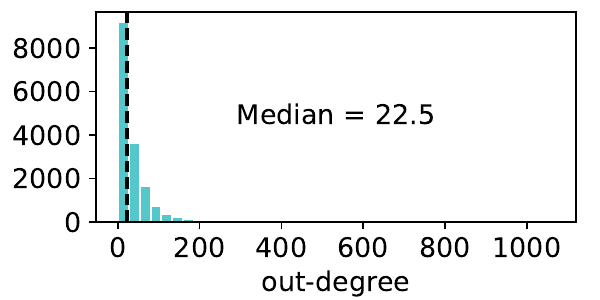}%\caption{fig1}
    }
    {\includegraphics[width=.24\textwidth]{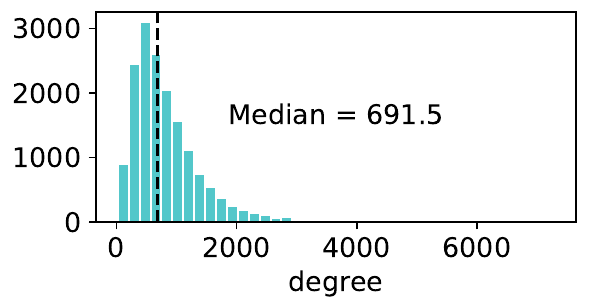}%\caption{fig1}
    }
    {\includegraphics[width=.24\textwidth]{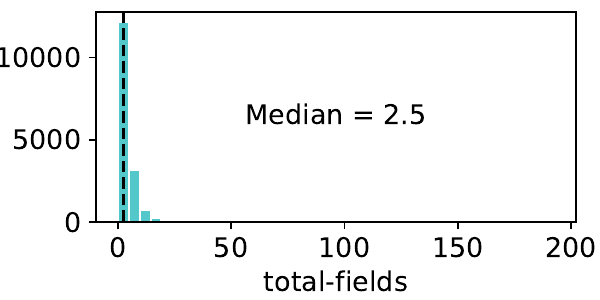}%\caption{fig1}
    }
    \quad\quad
    % \subfloat[The important features for InSep.]
    {\includegraphics[width=.24\textwidth]{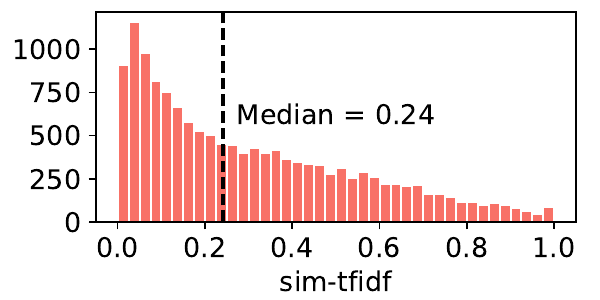}%\caption{fig1}
    }
    % \subfloat[The important features for InCol.]
    {\includegraphics[width=.24\textwidth]{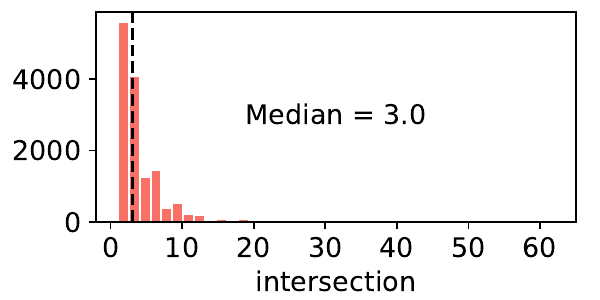}%\caption{fig1}
    }
    {\includegraphics[width=.24\textwidth]{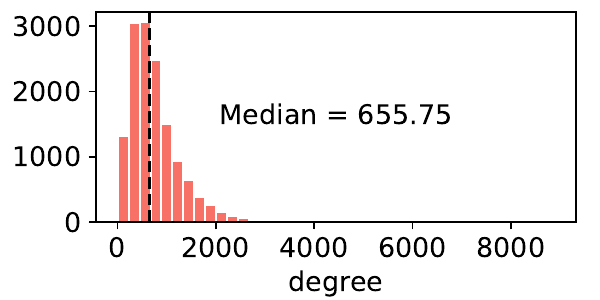}%\caption{fig1}
    }
    {\includegraphics[width=.24\textwidth]{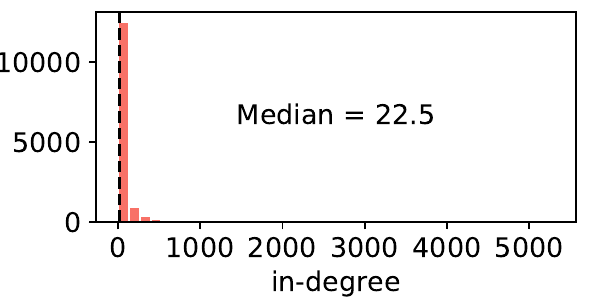}%\caption{fig1}
    }
    \caption{Distribution of the most important features for predicting InSep (Blue) and InCol (Red) for the 2nd experiment} 
    \label{fig: distributions_experiment2}
\end{figure*}

\begin{figure*}[t]
    \centering
    {\includegraphics[width=.24\textwidth]{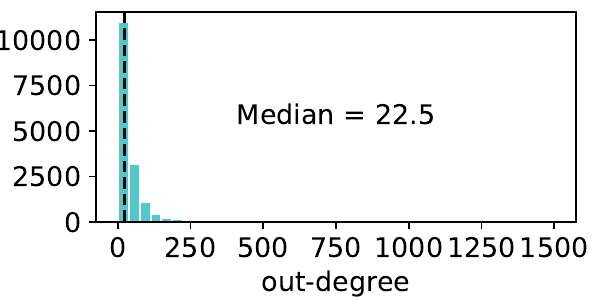}%\caption{fig1}
    }
    {\includegraphics[width=.24\textwidth]{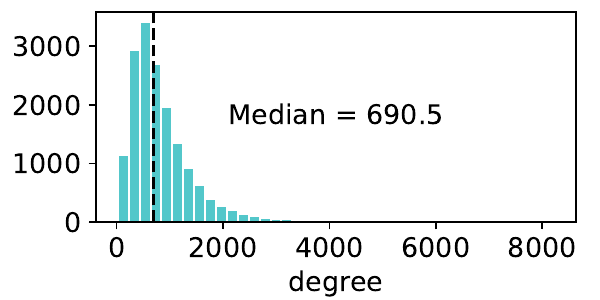}%\caption{fig1}
    }
    {\includegraphics[width=.24\textwidth]{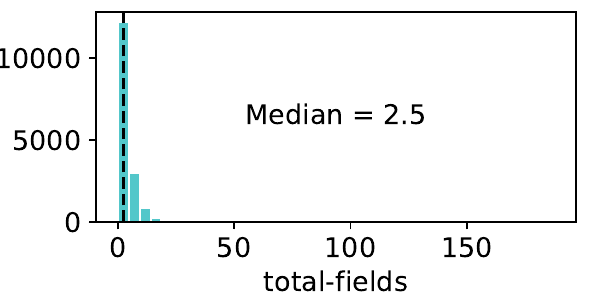}%\caption{fig1}
    }
    \quad\quad
    % \subfloat[The important features for InSep.]
    {\includegraphics[width=.24\textwidth]{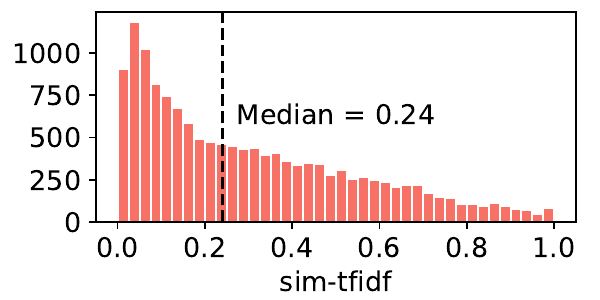}%\caption{fig1}
    }
    % \subfloat[The important features for InCol.]
    {\includegraphics[width=.24\textwidth]{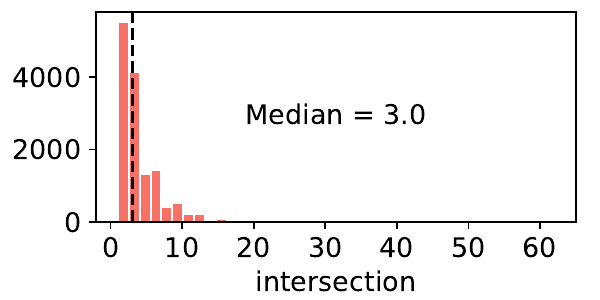}%\caption{fig1}
    }
    {\includegraphics[width=.24\textwidth]{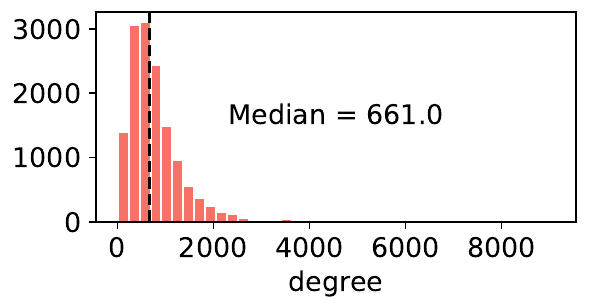}%\caption{fig1}
    }
    {\includegraphics[width=.24\textwidth]{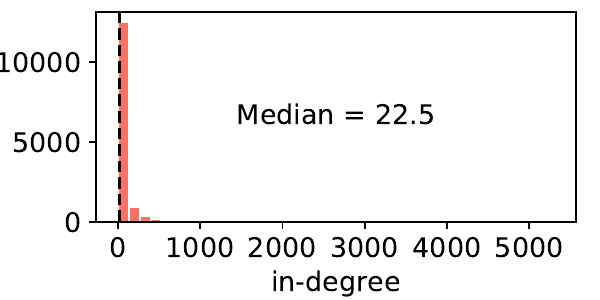}%\caption{fig1}
    }
    \caption{Distribution of the most important features for predicting InSep (Blue) and InCol (Red) for the 2nd experiment} 
    \label{fig: distributions_experiment3}
\end{figure*}

\change{R1.3}{Fig.~\ref{fig: distributions_experiment2} and Fig.~\ref{fig: distributions_experiment3} show the distributions of the features for predicting $\mathit{InSep}$ and $\mathit{InCol}$ for the 2nd and 3rd experiments. For half of the separated pairs, each entity in the pair, on average, has fewer than 22.5 out-going dependencies (in both experiments), shares fewer than 692 for the 2nd experiment (and 691 for the 3rd experiment) common terms with other entities, and implements no more than two fields.
On the other hand, the medians of \emph{sim-tfidf}, \emph{intersection}, \emph{degree}, and \emph{in-degree} among collocated pairs are 0.24, 3, 656, and 23 for the 2nd experiment (or 0.24, 3, 661, and 23 for the 3rd experiment) respectively.
The distributions of values for important features are highly skewed, which should be considered when analyzing the impact of features. 
Table~\ref{tab: differences between smells and non-smells_experiment2} and Table~\ref{tab: differences between smells and non-smells_experiment3}  further shows the median values of each important feature for smelly (\eg InSep) and non-smelly pairs (\eg Sep - InSep). 
We used the Mann-Whitney U test to test whether the distributions of each feature for smelly and non-smelly pairs are equal. 
As can be seen, all features show significant differences; however, only \emph{out-degree} and \emph{degree} in separated pairs, and \emph{in-degree} in collocated pairs, exhibit small effect sizes. }

\begin{table}
\caption{Medians for smelly and non-smelly pairs regarding the important pair characteristics (The 2nd Experiment)}
\label{tab: differences between smells and non-smells_experiment2}
\scriptsize
\centering
\begin{tabular}{rrr r@{\hspace{0.001cm}} rrr}
\toprule
 \multicolumn{3}{c}{\textbf{Separated Pairs}} &  &\multicolumn{3}{c}{\textbf{Collocated Pairs}} \\
 \cmidrule{1-3} \cmidrule{5-7}
\textbf{Char.} & \textbf{InSep} & \textbf{Sep. - InSep} &  & \textbf{Char.} &  \textbf{InCol} & \textbf{Col. - InCol}  \\
\midrule
\emph{out-degree} *** & \cellcolor[HTML]{D9D9D9}41	& \cellcolor[HTML]{D9D9D9}19   & & \emph{sim-tfidf} *** & 0.09	& 0.26  \\
\emph{degree}*** & \cellcolor[HTML]{D9D9D9}941.5	& \cellcolor[HTML]{D9D9D9}641.3   && \emph{intersection}*** & 2 & 3  \\
\emph{total-fields}*** & 3	& 2.5   && \emph{degree}* & 708 & 651 \\
-   & - &   -   &&  in-degree***   & \cellcolor[HTML]{D9D9D9}97.75  & \cellcolor[HTML]{D9D9D9}19.5  \\
% -   & - &   -   &&  comment-density*  & 0.79  & 0.67  \\
\bottomrule
\end{tabular}
\begin{tablenotes}
        \item * ``Sep.'' and ``Col.'' denote separated pairs and collocated pairs, respectively.
        \item * $p < 0.1$, $p < 0.01$, and $p < 0.001$ are denoted by *, **, and ***, respectively.
        \item * \colorbox{mygray}{Gray} results show small differences, while White results show an absolute difference smaller than 0.2.
\end{tablenotes}
\end{table}

\begin{table}
\caption{Medians for smelly and non-smelly pairs regarding the important pair characteristics (The 3rd Experiment)}
\label{tab: differences between smells and non-smells_experiment3}
\scriptsize
\centering
\begin{tabular}{rrr r@{\hspace{0.001cm}} rrr}
\toprule
 \multicolumn{3}{c}{\textbf{Separated Pairs}} &  &\multicolumn{3}{c}{\textbf{Collocated Pairs}} \\
 \cmidrule{1-3} \cmidrule{5-7}
\textbf{Char.} & \textbf{InSep} & \textbf{Sep. - InSep} &  & \textbf{Char.} &  \textbf{InCol} & \textbf{Col. - InCol}  \\
\midrule
\emph{out-degree} *** & \cellcolor[HTML]{D9D9D9}41.5	& \cellcolor[HTML]{D9D9D9}19.5   & & \emph{sim-tfidf} *** & 0.09	& 0.26  \\
\emph{degree}*** & \cellcolor[HTML]{D9D9D9}940.5	& \cellcolor[HTML]{D9D9D9}643   && \emph{intersection}*** & 2 & 3  \\
\emph{total-fields}*** & 3	& 2.5   && \emph{degree}*** & 705 & 657 \\
-   & - &   -   &&  in-degree***   & \cellcolor[HTML]{D9D9D9}98.75  & \cellcolor[HTML]{D9D9D9}20  \\
% -   & - &   -   &&  comment-density*  & 0.79  & 0.67  \\
\bottomrule
\end{tabular}
\begin{tablenotes}
        \item * ``Sep.'' and ``Col.'' denote separated pairs and collocated pairs, respectively.
        \item * $p < 0.1$, $p < 0.01$, and $p < 0.001$ are denoted by *, **, and ***, respectively.
        \item * \colorbox{mygray}{Gray} results show small differences, while White results show an absolute difference smaller than 0.2.
\end{tablenotes}
\end{table}

\begin{figure*}
    \centering
    {\includegraphics[width=.24\textwidth]{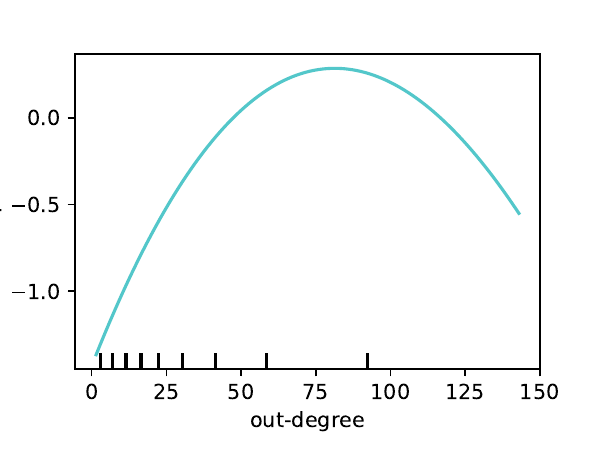}%\caption{fig1}
    }
    {\includegraphics[width=.24\textwidth]{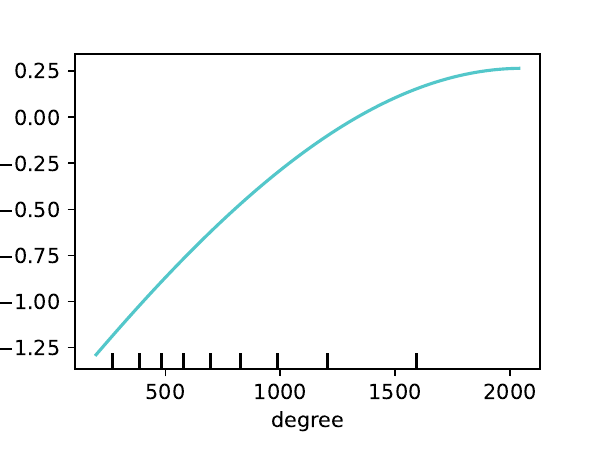}%\caption{fig1}
    }
    {\includegraphics[width=.24\textwidth]{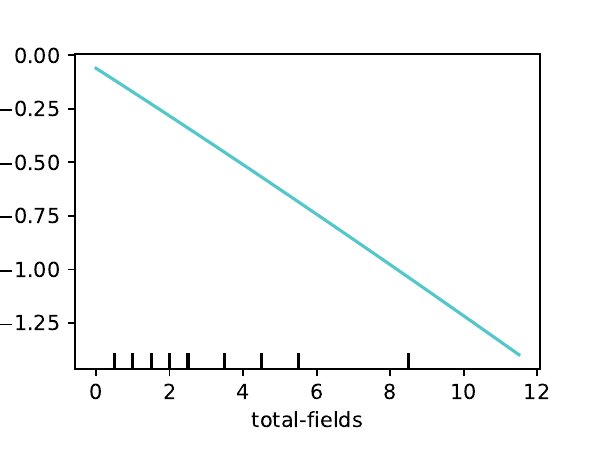}%\caption{fig1}
    }
    \quad
    \quad
    % \subfloat[The important features for InSep.]
    {\includegraphics[width=.24\textwidth]{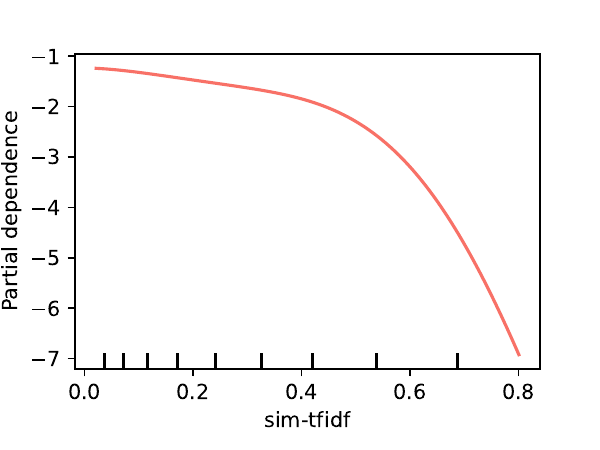}%\caption{fig1}
    }
    % \subfloat[The important features for InCol.]
    {\includegraphics[width=.24\textwidth]{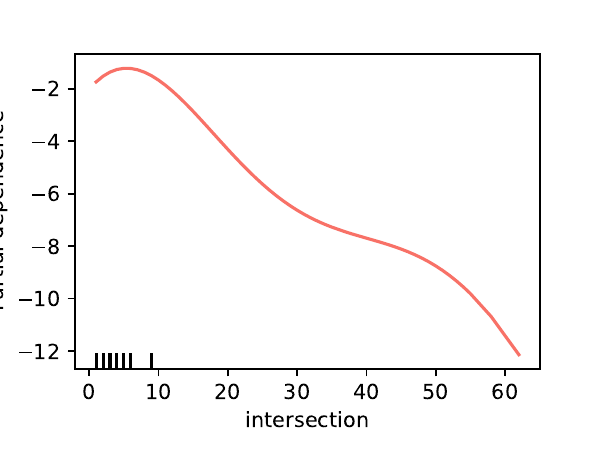}%\caption{fig1}
    }
    {\includegraphics[width=.24\textwidth]{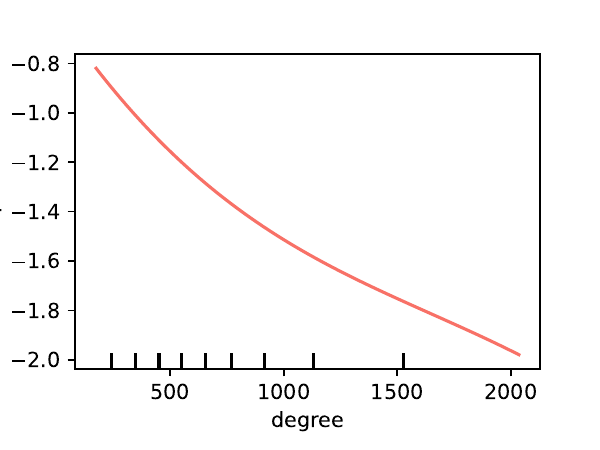}%\caption{fig1}
    }
    {\includegraphics[width=.24\textwidth]{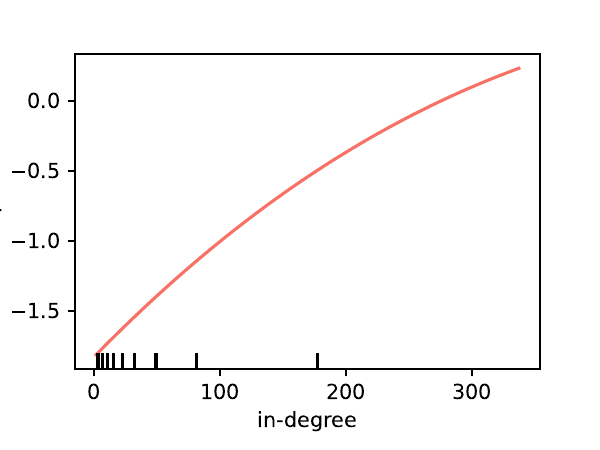}%\caption{fig1}
    }
    \caption{Partial dependence plots for each important feature for prediction InSep and InCol} 
    \label{fig: partial dependence_experiment2}
\end{figure*}

\begin{figure*}
    \centering
    {\includegraphics[width=.24\textwidth]{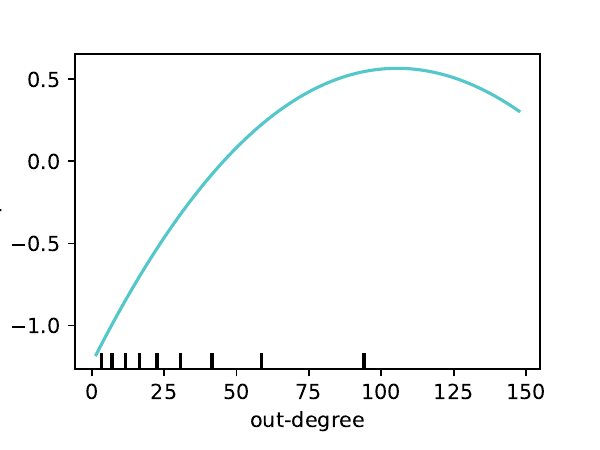}%\caption{fig1}
    }
    {\includegraphics[width=.24\textwidth]{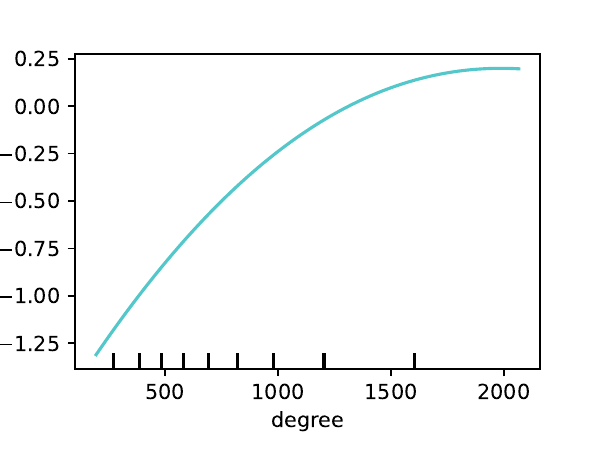}%\caption{fig1}
    }
    {\includegraphics[width=.24\textwidth]{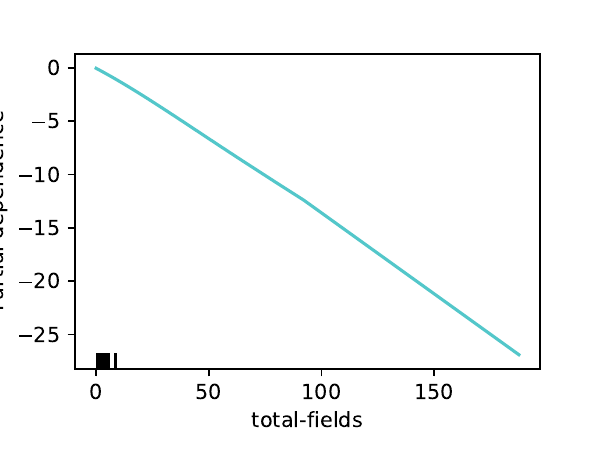}%\caption{fig1}
    }
    \quad
    \quad
    % \subfloat[The important features for InSep.]
    {\includegraphics[width=.24\textwidth]{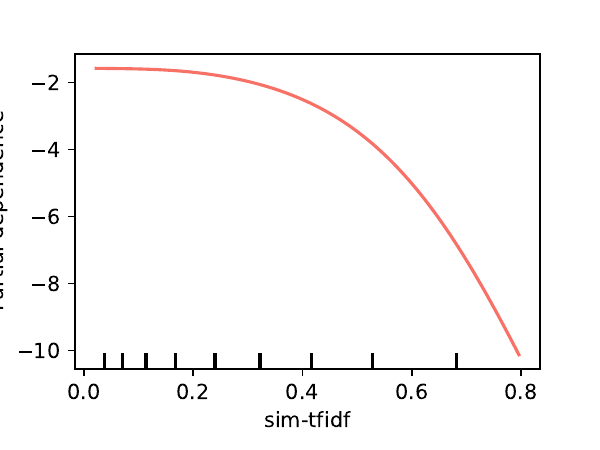}%\caption{fig1}
    }
    % \subfloat[The important features for InCol.]
    {\includegraphics[width=.24\textwidth]{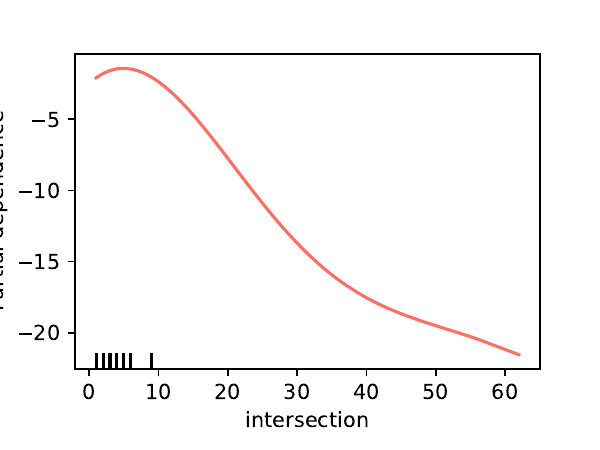}%\caption{fig1}
    }
    {\includegraphics[width=.24\textwidth]{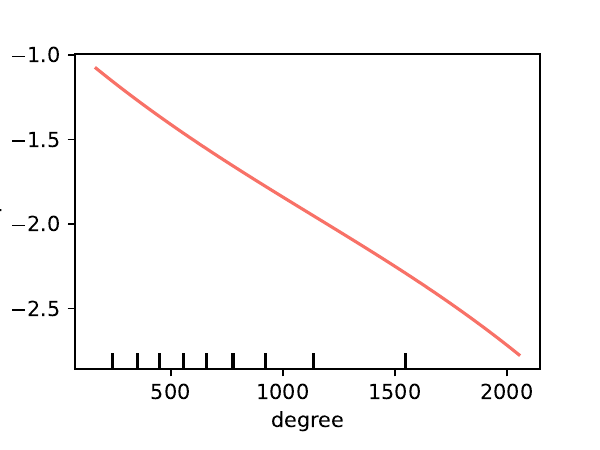}%\caption{fig1}
    }
    {\includegraphics[width=.24\textwidth]{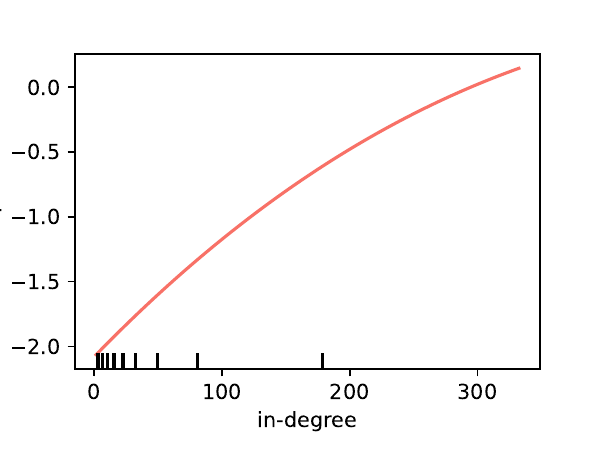}%\caption{fig1}
    }
    \caption{Partial dependence plots for each important feature for prediction InSep and InCol} 
    \label{fig: partial dependence_experiment3}
\end{figure*}

\change{}{To visualize how a change in a pair characteristic (feature) impacts the models' decision-making for each class, we draw \emph{Partial Dependence Plots (PDP)} for the most important features for the 2nd and 3rd experiments as presented in Fig.~\ref{fig: partial dependence_experiment2} and Fig.~\ref{fig: partial dependence_experiment3}. 
For separated pairs, we see that as the number of outgoing edges (dependencies on other entities) of a pair increases, the probability of predicting the pair as $\mathit{InSep}$ first increases and then drops.

For \emph{degree}, the partial dependence value in both Fig.~\ref{fig: partial dependence_experiment2} and Fig.~\ref{fig: partial dependence_experiment3} increases monotonically with higher \emph{degree} values.
In other words, separated pairs with a higher \emph{degree} value are more likely to be classified as $\mathit{InSep}$.
% This suggests that the more terms two entities share with other entities in the system, the more likely they are to be functionally relevant, thus they should be placed together rather than separated.
For \emph{total-fields}, the partial dependence decreases monotonically as the value increases -- separated pairs with fewer \emph{total-fields} are more likely to be classified as $\mathit{InSep}$.}
% In object-oriented design, classes with few fields are often relatively simple, which means that they are not central to a business functionality. 
% However, the number of fields seems not to be a strong predictor of $\mathit{InSep}$; its permutation importance is below 0.04 -- far lower than that of other two features in Fig.~\ref{fig: importance features} (a).
% Further inspection shows that low total fields only increases the likelihood of \emph{InSep} when combined with other evidence, \eg direct dependencies.
% As shown in Table~\ref{tab: differences between dependent and non-dependent separated pairs}, only directly dependent pairs with low field counts exhibit high \emph{InSep} predictions. 
% This suggests that when two dependent entities each have few fields, separating them may be unjustified due to insufficient encapsulated responsibilities. 

\change{}{For collocated pairs, we observe that higher values in \emph{sim-tfidf}, \emph{intersection}, and \emph{degree} generally decrease the likelihood that a collocated pair will be predicted as \emph{InCol}, while higher \emph{in-degree} increases it.
Specifically, as the \emph{sim-tfidf} value increases, the partial dependence decreases slightly initially and then drops significantly (after around 0.4).
That is, when the corresponding two entities have a low \emph{sim-tfidf} (\ie they are textually dissimilar), they are likely to be collocated inappropriately. 
% This kind of entity pairs actually dominates the dataset according to the distribution in Fig.~\ref{fig: distributions}.
However, two entities with a high \emph{sim-tfidf} may have implemented relevant functionalities, thus keeping them into one module follows the single responsibility principle (at the module level), reducing the probability of inducing \emph{InCol}.}

\change{}{Similarly, as the \emph{intersection} increases, the predicted probability of the corresponding pair as \emph{InCol} decreases monotonically (except a small peak in the beginning).
That is, as the two entities in a pair share more common terms, they are more likely to implement related responsibilities, making it less likely that their collocation is considered inappropriate.}

\change{}{For \emph{in-degree}, the partial dependence increases monotonically with a higher number of ingoing edges.
That is, the collocation of a pair with more ingoing edges is more likely to be identified as inappropriate (which leads to $\mathit{InCol}$).}
% According to Almugrin et al.~\cite{almugrin2016using}, more dependencies from others on an entity suggests more responsibilities it has in the system. 
% Therefore, when a pair has more ingoing edges, it implies that the corresponding entities are responsible for more functionalities. Collocating such entities could potentially overload the module, which may raise concerns about the suitability of the collocation. Libraries and util packages are an exception to this, as we discuss in more detail in Section \ref{sec: limitationsOfPairSmell}. 

